\documentclass[fleqn,usenatbib]{mnras}

\usepackage{newtxtext,newtxmath}
\usepackage{caption}
\usepackage{threeparttable}

\usepackage[T1]{fontenc}

\DeclareRobustCommand{\VAN}[3]{#2}
\let\VANthebibliography\thebibliography
\def\thebibliography{\DeclareRobustCommand{\VAN}[3]{##3}\VANthebibliography}

\usepackage{graphicx}	
\usepackage{amsmath}	

\title[Discovery of the shell structure via break radii in the outer halo of the Milky Way]{Discovery of the shell structure via break radii in the outer halo of the Milky Way}

\author[Dashuang Ye et al.]{
Dashuang Ye,$^{1}$
Cuihua Du,$^{1}$\thanks{E-mail: ducuihua@ucas.ac.cn}
Jianrong Shi$^{2,1}$
and Jun Ma$^{2,1}$
\\
$^{1}$College of Astronomy and Space Sciences, University of Chinese Academy of Sciences, Beijing 100049, P.R. China\\
$^{2}$Key Laboratory of Optical Astronomy, National Astronomical Observatories, Chinese Academy of Sciences, Beijing 100012, P.R.China\\
}

\date{Accepted XXX. Received YYY; in original form ZZZ}

\pubyear{2015}

\begin{document}
\label{firstpage}
\pagerange{\pageref{firstpage}--\pageref{lastpage}}
\maketitle

\begin{abstract}
Based on the \textit{Gaia} DR3 RR Lyrae catalog, we use two methods to fit the density profiles with an improved broken power law, and find that there are two break radii coinciding with the two apocenter pile-ups of high-eccentricity Gaia-Sausage-Enceladus (GSE) merger. 
Also, there is a break caused by the Sagittarius (Sgr) stream.
Combining the positions of all breaks, we briefly analyze the metallicity and its dispersion as a function of $r$ as well as its distribution in cylindrical coordinates.
For the clean sample, the $z\text{-to-}x$ ellipsoid axial ratio $q$ in $36\,{\rm kpc}\,\textless\,r\,\textless\,96\,{\rm kpc}$ becomes much smaller than that of the inner halo $(r\,\textless\,36\,{\rm kpc})$, while the major axis has a large uncertainty in the region of $36-66\,{\rm kpc}$ and the one in the region of $66-96\,{\rm kpc}$ is obviously different from that dominated by the Hercules-Aquila Cloud (HAC) and the Virgo Overdensity (VOD) in the inner halo, which indicates that there is an over-density structure distributed at low zenithal angles.
Finally, we found that the over-density structure in the outer halo ($r\,\textgreater\,50\,{\rm kpc}$) is shell-shaped and relatively metal-rich compared to the outer background halo.
We conclude that the shells could be the apocenter pile-ups of the high-eccentricity GSE merger, which is supported by previous numerical simulations. 
\end{abstract}




\section{Introduction}
Accounting for $\sim1 \%$ of the total stellar mass, the stellar halo 
plays a particularly crucial role in understanding the early formation of the Galaxy due to the dynamical and chemical fossil record that they preserve.   
Astronomers have long sought to constrain models for the formation and evolution of the Milky Way (MW) on the basis of the stellar and globular cluster populations and unrelaxed substructures that it contains. 
In the MW, the observed substructures, such as overdensities and clusters in integrals of motion \citep{helmi2018merger,myeong2019evidence,koppelman2019multiple,naidu2020evidence,malhan2022global}, indicate that the halo was likely built purely via mergers.  
A significant merger, i.e. the Sgr merger \citep{ibata1994dwarf,majewski2003two,belokurov2006field,hernitschek2017geometry,sesar2017100,li2019detecting,bellazzini2020globular}, generates the Sgr stream under the influence of tidal gravity, which can be divided into leading and trailing tails with distinct apocenters and pericenters, and the metallicity distribution of the trailing tail in south and north are significantly different \citep{belokurov2014precession}.
After the data release of the \textit{Gaia} satellite, our understanding of the Galactic formation history has been revolutionized.
One of the most insightful findings is a major merger of dwarf galaxy with the MW progenitor at redshift $z=1\sim2$ \citep[GSE,][]{belokurov2018co,helmi2018merger}.
The new insight brought by GSE is that the inner $(\textless\,30\,{\rm kpc})$ halo was largely built from one accretion event.
Members of GSE are the major constituents of the inner halo, 
they have high eccentricity, stongly radial velocity and near-zero rotation \citep{belokurov2018co}, and the distribution of their metallicities ([Fe/H]) has a peak ranging from -1.4 to -1.2 dex \citep{mackereth2020weighing,naidu2020evidence,das2020ages}.

However, GSE includes additionally many retrograde stars, which are actually derived from another merger called Sequoia \citep{myeong2019evidence}.
In addition to the above three significant accretion events, there are many other accretion events, such as Cetus \citep{newberg2009discovery,yuan2019revealing}, Pontus \citep{malhan2022global}, LMS-1/Wukong \citep{naidu2020evidence,yuan2020low,malhan2021evidence}, Arjuna \citep{naidu2020evidence}, I'itoi \citep{naidu2020evidence}, Thamnos 1 and 2 \citep{koppelman2019multiple}, Kraken/Koala \citep{kruijssen2020kraken,forbes2020reverse} and Helmi Streams \citep{helmi1999debris}. 
At the moment, the composition of distant ($r\,\textgreater\,40$\,kpc) halo is largely unknown due to low-quality astrometry and photometry dataset at such a large distance, although it is important for understanding the assembly history of the MW.

In general, the density profile of the Galactic stellar halo can reflect the accretion of the MW through apocentric pile-up. 
For example, a coincidence between the apocenter of large-eccentricity metal-rich halo stars (likely belonging to the GSE) and the stellar halo “break radius” at galactocentric distance $r\sim20$\,kpc was found by \citet{deason2018apocenter}.
Based on this fact, they suggest that the break resulted from the apocenter pile-up of high-eccentricity stars through a massive accretion event, which is exactly GSE subsequently identified by \citet{helmi2018merger} and \citet{belokurov2018co}.
So far, many models have been used to constrain the shape of the Galactic halo, such as single power-law \citep[SPL,][]{sesar2013exploring}, broken power-law \citep[BPL,][]{xue2015radial}, double power-law (DPL), cored power-law \citep[CPL,][]{iorio2018first} and Einasto \citep[EIN,][]{hernitschek2018profile} profiles.
In addition, a twice-broken power-law (TBPL) profile has also been adopted in several works \citep[e.g.][]{deason2014touching,xue2015radial,han2022stellar}.
A census of previous studies of the density profile was presented in the review by \citet{bland2016galaxy}. 
However, there are considerable spreads in the flattening ($q$) and the slope ($n$) for those best fitting models reported previously, for example, $q=0.2\longrightarrow0.8,n=-4.2$ \citep{xue2015radial}, $q=0.39\longrightarrow0.81,n=-4.7$ \citep{das2016characterizing}, $q=0.57\longrightarrow0.84,n=-2.96$ \citep{iorio2018first} and $q=0.5\longrightarrow1.0,n=-3.15$ \citep{miceli2008evidence}, which suggests that the SPL and BPL cannot reflect the true underlying distribution.

Despite these significant advancements, the density profile of the outer halo has not yet been well constrained due to a lack of adequate and accurate datasets. 
Various tracers such as blue horizontal branch stars \citep{deason2011milky,das2016characterizing}, K giants \citep{xue2015radial} and main sequence turnoff stars \citep{juric2008milky}, have been used to study the radial density profile. 
Pulsating horizontal branch stars, also known as RR Lyrae (RRL), the majority of them are metal-poor and old (age$\,\textgreater\,9,10\,$Gyr) with low mass ($0.6-0.8\,\rm M_{\odot}$), their radial velocities are biased by a range of $\rm 40-70\,km\,s^{-1}$ \citep{liu1991synthetic,drake2013probing}, because their surfaces contract or expand regularly with periods shorter than a day.
RRL serve as a standard candle to measure distance \citep{muraveva2018rr,li2022photometric}, and is a intrinsically bright tracer of the Galactic halo \citep{sesar2013exploring,sesar2017100,hernitschek2017geometry,iorio2018first,hernitschek2018profile,iorio2019shape,ramos2020full,iorio2021chemo}.
The \textit{Gaia} DR3 \citep{vallenari2022gaia} from the \textit{Gaia} space observatory \citep{prusti2016gaia} provides a large sample of 270,905 RRL with high completeness, including 174,947 RRab, 93,952 RRc and 2,006 RRd \citep{clementini2022gaia}, based on this dataset, we, therefore, study the accumulate history of halo by analyzing the fitting results of the density profile and metallicity.

It is well known that stars on very large circular orbits prefer constant velocities, while those stars on extremely flat elliptical orbits move rapidly through their points of closest approach and slow down at their furthest extent. 
The inevitable slow-down leads to a build-up of stars at apocenter.
With the aid of the simulation in the context of the cold dark matter (CDM) model, \citet{cooper2011formation} found that the disruption of a satellite system on a near-radial orbit can create the observed ``shell'' structures in the halo of the nearby galaxy NGC 7600.
The two apocenter pile-ups, corresponding to a `double-break' at 15-18\,kpc and 28-30\,kpc, are attributed to the Hercules-Aquila Cloud \citep[HAC,][]{belokurov2007} and the Virgo Overdensity \citep[VOD,][]{vivas2001}, which are both from the GSE merger \citep{perottoni2022}.
Furthermore, based on the radial density and angular momentum distribution, \cite{naidu2021reconstructing} applied N-body galaxy simulations of the merger similar to GSE to study its accretion with the MW, and predicted that there are two breaks at $15\text{-}18~{\rm kpc}$ and $30~$kpc using TBPL to fit the density profile, which is coincident with the final two apocenters of the GSE before it fully merged with the MW. 

In this paper, we explore the shell-shaped structure by fitting the density profile of RRab sample with precise distances using the improved broken power law.
In Section \ref{Methods}, we describe the selection of two initial samples, the new broken density profile and two fitting methods, namely goodness-of-fit (GFFM) and classical fitting methods as well as how to get the candidates of shells. 
We analyze the best-fitting results and discuss the metallicity distribution and its dispersion as a function of radius in Section \ref{Results}.
The conclusion and summary are presented in Section \ref{Conclusions and discussion}.

\section{Methods}\label{Methods}

\subsection{Dataset and selection criteria}\label{Dataset and selection criteria}

\begin{figure}
	\includegraphics[width=\columnwidth]{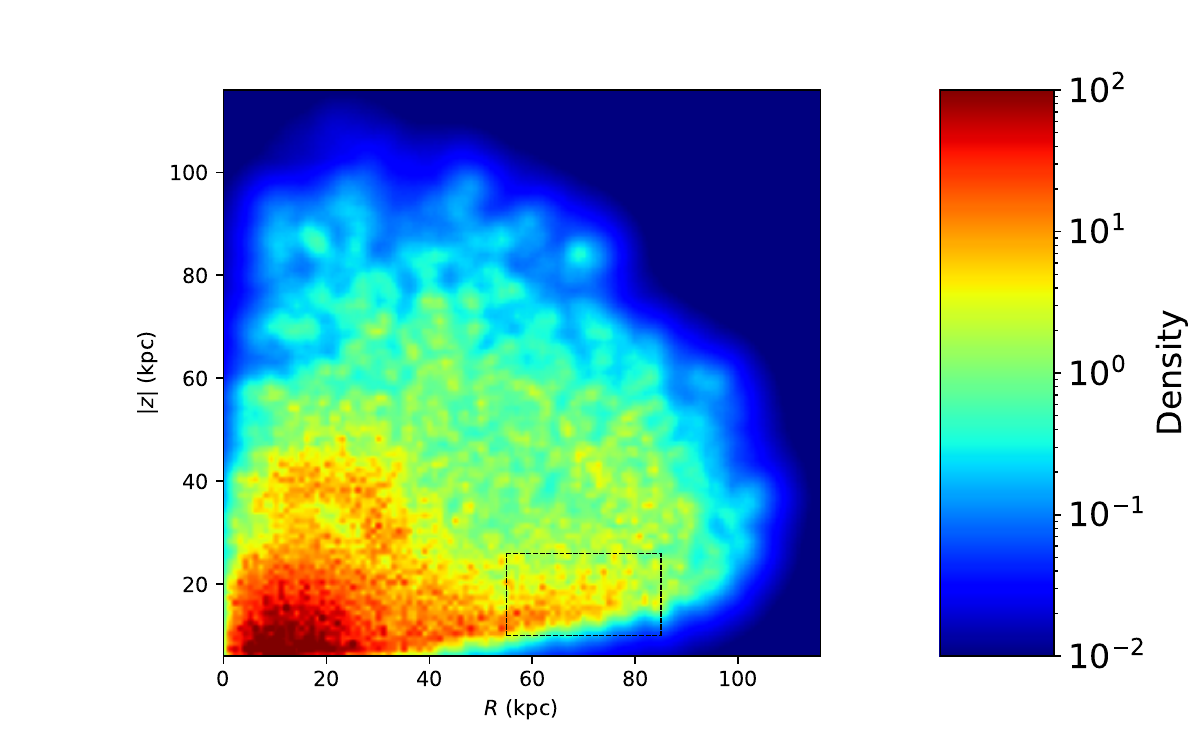}
    \includegraphics[width=\columnwidth]{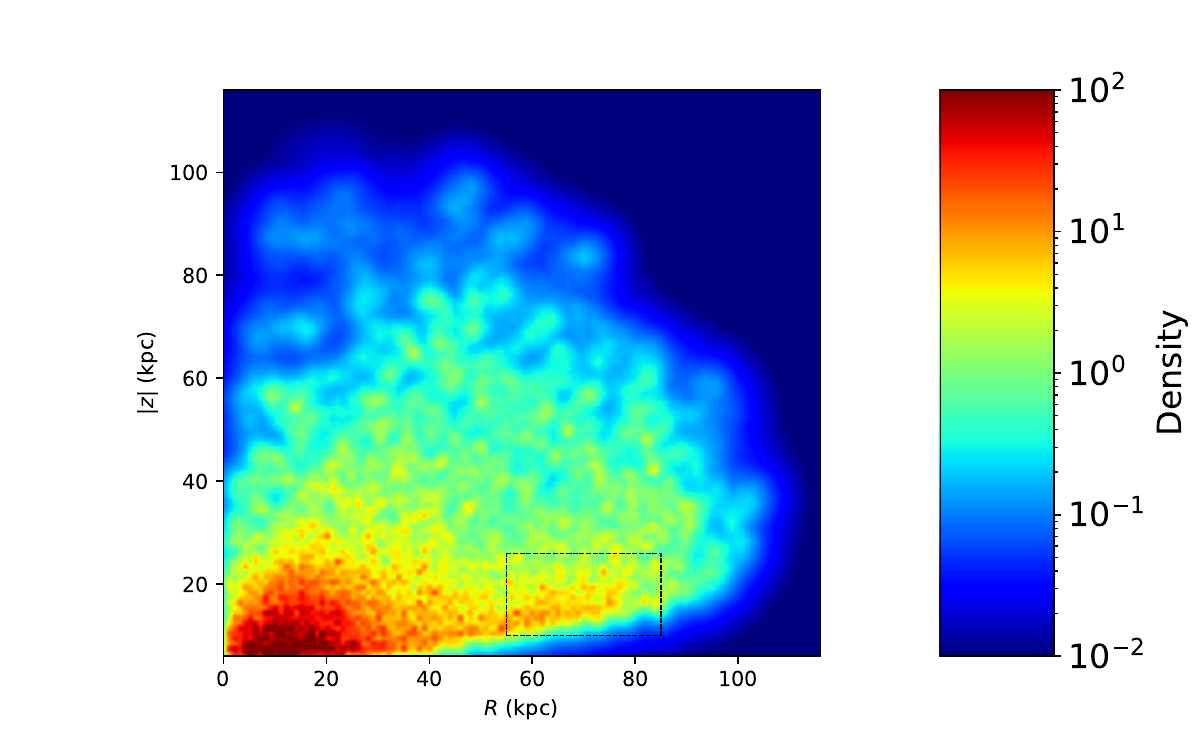}
    \caption{The $R-|z|$ distributions of our total sample (top) and clean sample (bottom). 
    The color bar on the right side indicates the number density.}\label{figR_zall}
\end{figure}

In this study, we adopt the left-handed Galactocentric reference frame \citep{iorio2018first}.
The Cartesian coordinates are indicated by $x$, $y$, and $z$; 
the cylindrical and spherical radii are $R$ and $r$, respectively; the zenithal and azimuthal angles are described by $\theta$, $\phi$, and the Galactic longitude and latitude are adopted as $l$ and $b$, respectively.
The Sun is located at $x_{\odot}=R_{\odot}=8.13\,$kpc \citep{abuter2018detection}, and $z_{\odot} = 0\,$kpc \citep{iorio2018first}.
We adopt the Sun's  peculiar motion as $ \left( U_{\odot},V_{\odot},W_{\odot} \right) = \left(-11.10\pm1.23,12.24\pm2.05,7.25\pm0.63 \right)\,$km\,s$^{-1} $ \citep{schonrich2010local}, while the local standard of rest (LSR) velocity as $V_{\rm LSR} = 238\pm9\,$km\,s$^{-1} $ \citep{schonrich2012galactic}.

Here, only the RRab stars have been considered due to their good completeness.
Since various checks on the derived photometric metallicities and distances have been done for the globular clusters and Magellanic Clouds \citep[e.g.][]{li2022photometric}, they are sufficiently accurate to investigate structures in the stellar halo.
We use their method to calculate the metallicities and 3D positions of our sample stars, but instead, we apply the selection criteria related to the proper motions and positions of the Magellanic Clouds in \citet{iorio2019shape} to select our member stars and correct the \textit{Gaia} G-magnitudes due to dust reddening. 
The median error of our distances is $0.068^{+0.016}_{-0.006}$, and the distances and metallicities of the Magellanic Clouds are consistent with those presented by \citet{li2022photometric}.
Derivation of metallicity requires the fundamental period $P$ and the phase difference between the third and the first harmonics of the light-curve decomposition $\Phi_{31}$, and the [Fe/H]-$P$-$\Phi_{31}$ relation for RRab stars \citep{li2022photometric} is expressed in equation (1).
The absolute magnitudes ($M_{\rm G}$) can be estimated using the $M_{\rm G}-{\rm [Fe/H]}$ relation in \citet{li2022photometric}, namely equation (2).
\citet{iorio2019shape} and \citet{Vasiliev2021tango} assumed absolute magnitudes to be constant when calculating distances.
\citet{ramos2020full} assignd an average metallicity -1.61\,dex to those stars without available metallicities.
In our study, for $\sim60,000$ stars without $\Phi_{31}$, these values are drawn from their overall 2D distribution considering the whole \textit{Gaia} catalogue, similar to that described in \citet{iorio2021chemo}, using the extreme-deconvolution algorithm \citep[XDGMM,][]{bovy2015}.
\begin{equation}
\begin{aligned}
    {\rm[Fe/H]} =& (2.669\pm0.039) + (-8.641\pm0.076)\times P\\ 
    +& (-0.907\pm0.022)\times\Phi_{31}+ (0.794\pm0.034)\times P\times\Phi_{31}\\ 
    +& (0.363\pm0.005)\times(\Phi_{31})^{2},
\end{aligned}
\end{equation}
\begin{equation}
    M_{\rm G} = (0.254\pm0.009){\rm [Fe/H]}+(1.002\pm0.012).
\end{equation}
In this work, the turning points of metallicity derived from these $\sim60,000$ stars without $\Phi_{31}$ are included in those turning points from stars with $\Phi_{31}$.
It needs to be noted that the mock metallicities ($\rm [Fe/H]_{mean}\sim-1.43$) are generally higher than those derived from $P$ and $\Phi_{31}$ ($\rm [Fe/H]_{mean}\sim-1.64$), however, this does not affect the accuracy of distance.
We make the mock metallicities close to the fitting curve between the metallicity and $r$, which only results in a relative change of $0.0239^{+0.0149}_{-0.0089}$ for distance, so the distances derived from the mock metallicities are accurate enough to explore all obvious break radii caused by apocenter pile-ups.

Regions with high dust extinction can bring some uncertainties in the study of the stellar distribution in the Galaxy.
The spatial cuts to exclude the Galactic disc and the regions with poor completeness and high dust extinction at low latitudes are as follows:\newline
\begin{equation}
S_{b} = \left\{
\begin{aligned}
0, & & {\lvert b \rvert\,\textless\,10^{\circ}}\\
1, & & {\rm else.}\\
\end{aligned}
\right.
\end{equation}
\newline
\begin{equation}
 S_{z} = \left\{
\begin{aligned}
0, & & {\lvert z \rvert\,\textless\,6\,\rm{kpc}}\\
1, & & {\rm else.}\\
\end{aligned}
\right.
\end{equation}
We apply the selection criteria from \citet{iorio2021chemo} to remove the most obvious compact structures, including artifacts/contaminants, globular clusters, dwarf satellites \citep{mateu2018fourteen}, the Sgr dwarf (not including the Sgr stream), Large Magellanic Cloud (LMC) and Small Magellanic Cloud (SMC), but we additionally consider the selection criteria for globular clusters: 
\newline
\begin{equation}
\left\{
\begin{array}{l}
\lvert \mu_{\alpha} - \mu_{\alpha,{\rm gc}} \rvert \,\textless \,10 \sigma_{\mu_{\alpha,{\rm gc}}}, \\
\lvert \mu_{\delta} - \mu_{\delta,{\rm gc}} \rvert \,\textless \,10 \sigma_{\mu_{\delta,{\rm gc}}}. \\
\end{array}
\right.
\end{equation}
\newline
where $\mu_{\alpha,{\rm gc}}$, $\mu_{\delta,{\rm gc}}$ are the proper motions, and $\sigma_{\mu_{\alpha,{\rm gc}}}$, $\sigma_{\mu_{\delta,{\rm gc}}}$ are their errors, respectively.
The spatial positions, tidal radii and half-light radii of the globular clusters used in this work come from the Harris catalogue\footnote{\url{https://physics.mcmaster.ca/~harris/Databases.html}} \citep{harris1996catalog}, while their proper motions and uncertainties are taken from the catalogue\footnote{\url{https://github.com/GalacticDynamics-Oxford/GaiaTools}} from \citet{vasiliev2021gaia}.

In order to excise the globular clusters without \textit{Gaia} proper motions \citep{vasiliev2021gaia} and all dwarf galaxies \citep{mateu2018fourteen}, the selection function is as follows:
\begin{equation}
S_{\rm dwgc} = \left\{
\begin{aligned}
0, & & {\rm spatial\ conditions\ (dwgc)}\\
1, & & {\rm else.}\\
\end{aligned}
\right. 
\end{equation}
For the globular clusters with proper motions, the selection function is given as:
\begin{equation}
S_{\rm gc} = \left\{
\begin{aligned}
1 - f_{\rm gc}, & & {\rm spatial\ conditions\ (gc)}\\
1, & & {\rm else.}\\
\end{aligned}
\right. 
\end{equation}
\newline
where $ f_{\rm gc} $ is the number-ratio of selected stars of all conditions versus only under spatial conditions for each globular cluster.
The selection function to cut the Sgr dwarf galaxy is given as:\newline
\begin{equation}
S_{\rm Sgr} = \left\{
\begin{aligned}
1 - f_{\rm Sgr}, & & {\rm spatial\ conditions\ (Sgr)}\\
1, & & {\rm else.}\\
\end{aligned}
\right.
\end{equation}
\newline
where $ f_{\rm Sgr} $ is the ratio of selected stars considering all conditions versus only under spatial conditions in each 2D $ \left( \widetilde{B} - \widetilde{B}_{\rm Sgr} \right) $ versus $ \left( \widetilde{\Lambda} - \widetilde{\Lambda}_{\rm Sgr} \right) $ bin ($1^{\circ}\times 1^{\circ}$) within $\lvert \widetilde{B} - \widetilde{B}_{\rm Sgr} \rvert\,\textless\,9^{\circ}$ and $\lvert \widetilde{\Lambda} - \widetilde{\Lambda}_{\rm Sgr} \rvert\,\textless\,50^{\circ}$, where $\widetilde{B}$ and $\widetilde{\Lambda}$ are the latitude and longitude in the coordinate system aligned with the Sgr stream as defined in \citet{belokurov2014precession}, and $\widetilde{B}_{\rm Sgr}=4.24^{\circ}$ and $\widetilde{\Lambda}_{\rm Sgr}=-1.55^{\circ}$ represent the position of the Sgr dwarf.
The selection function of outer LMC and SMC is given as:\newline
\begin{equation}
S_{\rm outer\ LMC,SMC} = \left\{
\begin{aligned}
1 - f_{\rm LMC}, & & {\rm spatial\ conditions\ (outer\ LMC)}\\
1 - f_{\rm SMC}, & & {\rm spatial\ conditions\ (outer\ SMC)}\\
1, & & {\rm else.}\\
\end{aligned}
\right. 
\end{equation}
\newline
where, $ f_{\rm LMC} $ is the ratio of selected stars considering all conditions versus only under spatial conditions in each 1D bin ($1^{\circ}$) within $5^{\circ}\,\textless\,\eta_{\rm{LMC}}\,\textless\,16^{\circ}$ ($\eta_{\rm{LMC}}$ is the angle with the LMC core),
and $f_{\rm SMC}$ is the similar ratio in each 1D bin ($1^{\circ}$) within $5^{\circ}\,\textless\,\eta_{\rm{SMC}}\,\textless\,12^{\circ}$ ($\eta_{\rm{SMC}}$ is the angle with the SMC core). 
Note that the spatial conditions (outer SMC) denote that the spatial region of outer SMC minus its intersection with that of outer LMC.
The selection function to excise inner LMC and SMC is given as:\newline
\begin{equation}
S_{\rm inner\ LMC,SMC} = \left\{
\begin{aligned}
0, & & {\rm spatial\ conditions\ (inner\ LMC)}\\
0, & & {\rm spatial\ conditions\ (inner\ SMC)} \\
1, & & {\rm else.}\\
\end{aligned}
\right. 
\end{equation}
After the aforementioned selections, we obtain 32,019 stars (31,707 RRab stars with the \textit{Gaia} proper motions) within $116\,{\rm kpc}$ from the Galactic centre as our total sample, as shown in the top panel of Figure \ref{figR_zall}, which will be used in the fitting based on goodness-of-fit method (see Section \ref{GFFM}).

In order to obtain a complete sample, we remove globular clusters, Sgr dwarf galaxy, SMC and LMC not only with spatial cuts but also with proper motion and G-magnitude. 
However, some breaks in density profile may be induced by proportional reductions in their selection functions.
In order to rule out these potential possibilities, we only use spatial cuts to remove them.
Therefore, for the globular clusters, SMC and LMC, the corresponding number-ratio of selected stars $f$ in their selection functions will be set to 1, and the selection function of the Sgr dwarf galaxy and stream is as follows:\newline
\begin{equation}
S_{\rm Sgr} = \left\{
\begin{aligned}
0, & & {\lvert \widetilde{B}\rvert\,\textless\,11^{\circ}\ {\rm and}\ D_{\rm sun}\,\textgreater\,15\,{\rm kpc}}\\
1, & & {\rm else.}\\
\end{aligned}
\right. 
\end{equation}
Keeping other selection conditions unchanged, we finally get 22,969 RRab stars as our clean sample, as shown in the bottom panel of Figure \ref{figR_zall}, which will be used to fit the density profile by the classical fitting method (see Section \ref{Classical fitting method}).

\subsection{Broken power law}\label{Broken power law}

\begin{figure*}
	\includegraphics[width=\textwidth]{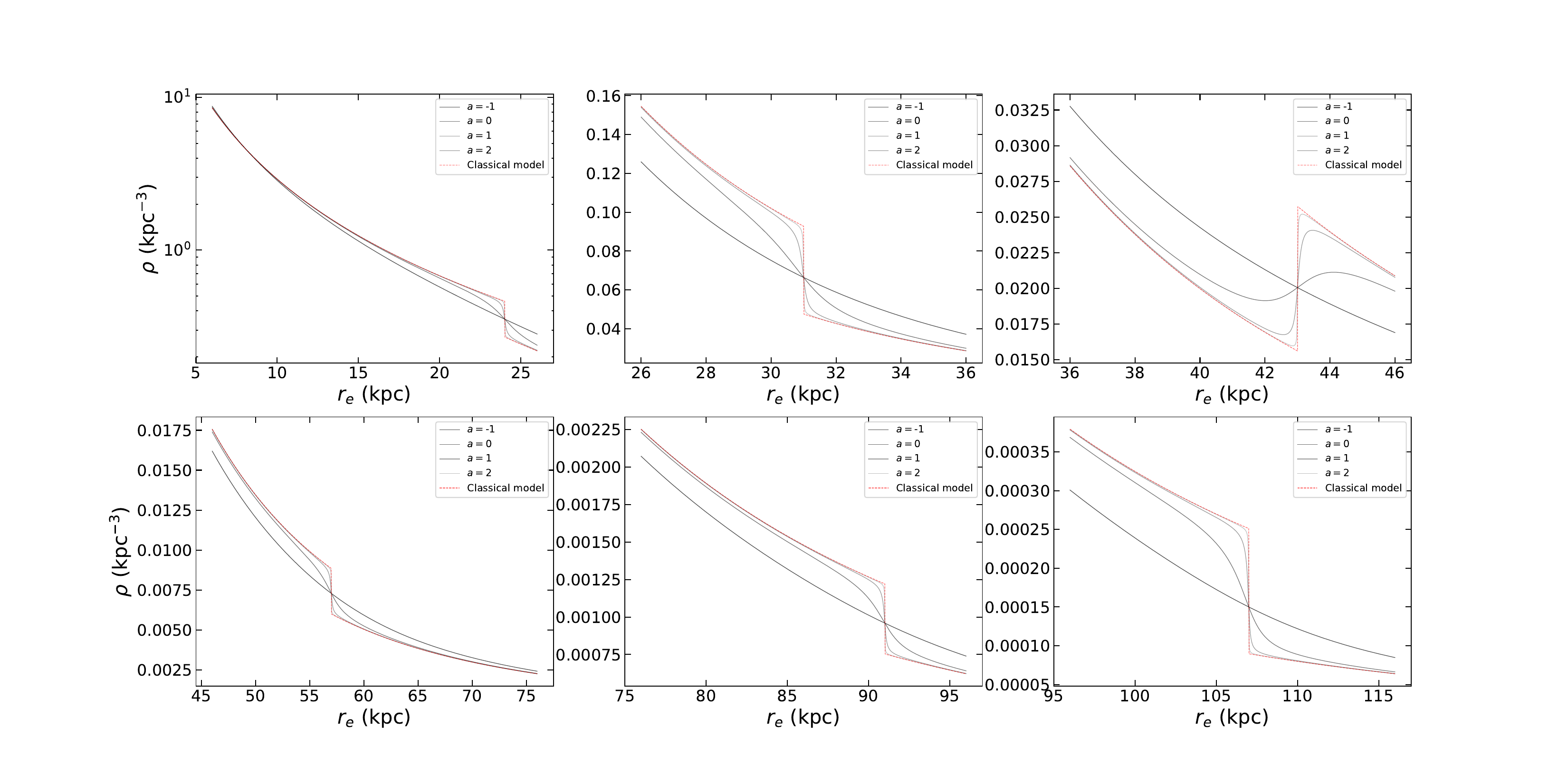}
    \caption{Classical BPL density profile (red dashed line) and SBPL density profile with various $a$ (black solid lines with various degrees of transparency) as a function of the elliptical radius $r_{\rm e}$. Among them, the parameters we substitute into the SBPL model are listed in Table \ref{table_result}.}\label{rho_re}
\end{figure*}

The break scale can reflect the spatial characteristics of the substructure edge, for example, the break scale of diffuse substructure is larger than that of compact one.
Moreover, in order to identify false breaks with an unreasonable scale, such as a large-scale break due to the observation limitations, we introduce a new broken power law in which a specific parameter indicates the break scale:
\newline
\begin{equation}
\rho_{\rm halo} = \rho_{\odot}^{\rm RRab}\left( \frac{R_{\odot}}{r_{\mathrm{e}}} \right)^{\left (n + \sum\limits_{i=0}^{m} (0.5\delta n_{i} +\left( \frac{\delta n_{i}}{\pi} \right){\rm arctan}\left((10^{a_{i}})\times\left(r_{\mathrm{e}}-r_{\mathrm{break},i}\right)\right))   \right) }. 
\end{equation}
\newline
where $ \rho_{\odot}^{\rm RRab}=4.5\,{\rm kpc}^{-3}$ \citep{sesar2013exploring} is the number density of RRab stars in the vicinity of the Sun.
The elliptical radius is defined as $r_{\rm e}=\sqrt{\hat{x}^{2}+(\hat{y}/p)^{2}+\left(\hat{z}/q\right)^{2}}$, here $ p $ and $ q $ are, respectively, the $y$-to-$x$ and $z$-to-$x$ ellipsoid axial ratios ($\hat{\textbf{r}} = \textbf{R}_{y}(\beta)\textbf{R}_{x}(\eta)\textbf{R}_{z}(\gamma)\textbf{r}$, $\textbf{R}_{y}(\beta)$, $\textbf{R}_{x}(\eta)$ and $\textbf{R}_{z}(\gamma)$ are counterclockwise rotation matrices around the $y\text{-axis, }x\text{-axis and }z\text{-axis}$, respectively, so $\gamma$ and $\eta$ determine the direction of the major axis.) .
$r_{\mathrm{break},i}$ and $\delta n_{i}$ represent the radius and slope variation of the $i$-th break.
$a_i$ is a parameter associated with the scale of the $i$-th break and decreases with the corresponding break scale.
Figure \ref{rho_re} shows $\rho$ as a function of $r_{\rm e}$ for various values of $a$, where all breaks are obtained by GFFM.
Considering that our aim is to explore shells or rings formed by the mergers \citep[e.g.][]{pop2018formation,karademir2019outer}, we will not introduce the other density profiles here.

\subsection{Fitting method based on goodness-of-fit}\label{GFFM}

We divide our total sample into the following nine bins according to $r$: $(6,16)$\,kpc, $(16,26)$\,kpc, $(26,36)$\,kpc, $(36,46)$\,kpc, $(46,56)$\,kpc, $(56,76)$\,kpc, $(66,86)$\,kpc, $(76,96)$\,kpc, $(96,116)$\,kpc.
With fixed $a,p,\gamma,\eta$ and $\beta$ of $(0,1,0,0,0)$ in this method, the computation time will be greatly reduced, while a shell-shaped structure can still be detected.
We fit the data in each bin with two density laws: single power law (SPL) and singly broken power law (SBPL), and then compare their maximum goodness-of-fits $(\rm GF_{max})$.
If the $\rm GF_{max}$ of SBPL is much larger than that of SPL, we consider the break to be justified, otherwise, we merge the interval with its neighbor to fit the SPL and SBPL models again.
Within 56\,kpc from the Galactic centre the size of each bin is 10\,kpc, while it is 20\,kpc beyond 56\,kpc for the following reasons:
(\romannumeral1) the apocenters of GSE and the Sgr leading stream locate within 56\,kpc, therefore, it can avoid missing breaks in larger bins;
(\romannumeral2) within $ r_{\rm break}\pm5$\,kpc, 3.08\,kpc, 1.96\,kpc, 1.38\,kpc, 1\,kpc, $n$ will increase by 0.874$\delta n$, 0.8$\delta n$, 0.7$\delta n$, 0.6$\delta n$, 0.5$\delta n$, respectively, so that each bin is large enough to include the core of break;
(\romannumeral3) previous studies have shown that the flattening $q$ becomes larger and more stable as $r$ increases;
(\romannumeral4) if fitting the data in a larger interval such as $(6,36)\,$kpc, we have to use SPL, SBPL, TBPL or broken power-law model with three breaks and compare their $\rm GF_{\rm max}$, in which each additional break will make our computation time increases exponentially;
(\romannumeral5) if fitting the data in a smaller interval, more spurious breaks could be detected and the statistic goodness-of-fit (GF) will be less reliable because it is derived from local information.

We apply the kernel density estimation \citep[KDE, the kernel is a `gaussian', and the optimal bandwidth 0.886 is given by the cross-validation method,][]{weiss1991computer} in the Python module \textbf{sci-kit learn} to estimate the radial density at some radii (for example 6.5\,kpc, 7.5\,kpc, 8.5\,kpc, 9.5\,kpc, 10.5\,kpc, 11.5\,kpc, 12.5\,kpc, 13.5\,kpc, 14.5\,kpc, 15.5\,kpc in 6-16\,kpc).
Furthermore, we need to construct a sufficiently dense grid of parameters to avoid missing the best-fitting value for the sensitive parameters.
Considering that $q\ (\textless1)$ increases with $r$ and $n_{2}\approx n_{1}+\delta n_{1}$ ($n_{2}$ is the best-fitting slope in next bin), we can narrow the range of grid and increase its density (see Table \ref{table1}).

We use SBPL to fit the dataset for each bin according to the following steps:
(\romannumeral1) substituting each point into the SBPL density profile;
(\romannumeral2) calculating the integral of SBPL density profile for each sub-interval, such as $ 6.5\pm 0.5 $\,kpc, as the model value while considering all the selection functions except that of the Sgr stream;
(\romannumeral3) calculating the GF based on both model values and observations from the KDE.
GF is defined as follows:
\begin{equation}
{\rm GF} \equiv 1-\frac{\sum\limits_{i=1}^{n}\left(y_{{\rm model},i}-y_{i}\right)^{2}}{\sum\limits_{i=1}^{n}\left(y_{i}-\overline{y}\right)^{2}}. 
\end{equation}
where $ y_{{\rm model},i} $ and $ y_{i} $, respectively, indicate the model values and observations, $\overline{y}$ is the mean observation value, and $n$ is the number of observations.
The approach for fitting the data within each interval using SPL is analogous to the aforementioned steps.

As listed in Table \ref{table2}, there is negligible difference in high $\rm GF_{max}$ ($\textgreater0.9$) between SPL and SBPL density profiles in the bins of $(6,16)\,{\rm kpc}$ and $(56,76)\,{\rm kpc}$ ($\rm\Delta GF_{max}=-0.002,-0.021$).
The distribution of sources in the two bins is equally fit well by SPL and SBPL, and the mode of $\delta n$, as listed in Table \ref{table1}, is approximately 0.1, so the breaks at $r_{\rm e}\sim10\,{\rm kpc}$ and $\sim60\,{\rm kpc}$ could be insignificant.
We combine the two bins with their adjacent bins into two larger bins, namely $(6,26)\,{\rm kpc}$ and $(46,76)\,{\rm kpc}$, respectively, and then use SPL and SBPL to fit the data in each larger bin.
We conclude that the break at 10\,kpc does not exist, and the break at 60\,kpc is insignificant due to its proximity to the inner boundary of the bin.

\begin{figure}
	\includegraphics[width=\columnwidth]{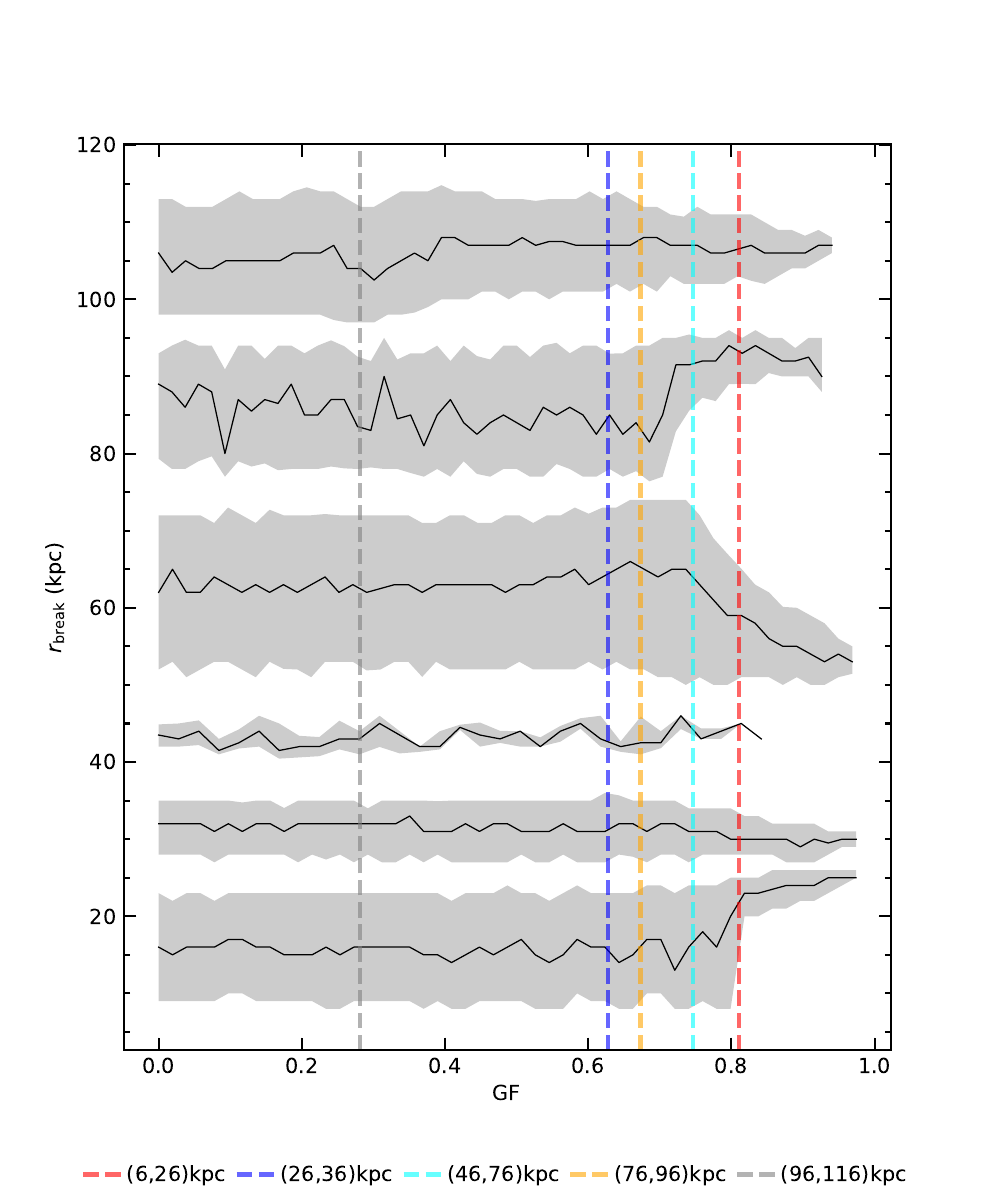}
    \caption{Break radius $r_{\rm break}$ as a function of Goodness of fit and its uncertainty (gray background).
    Each dashed line represents the highest goodness-of-fit $(\rm GF_{max})$ of SPL in each bin, which is listed in Table \ref{table_result} as the threshold of goodness-of-fit $(\rm GF_t)$.
    Note that the $\rm GF_{max}$ value of SPL in the bin of $36-46\,{\rm kpc}$ is negative and therefore not used.}\label{rbreak_GF}
\end{figure}

\begin{figure}
	\includegraphics[width=\columnwidth]{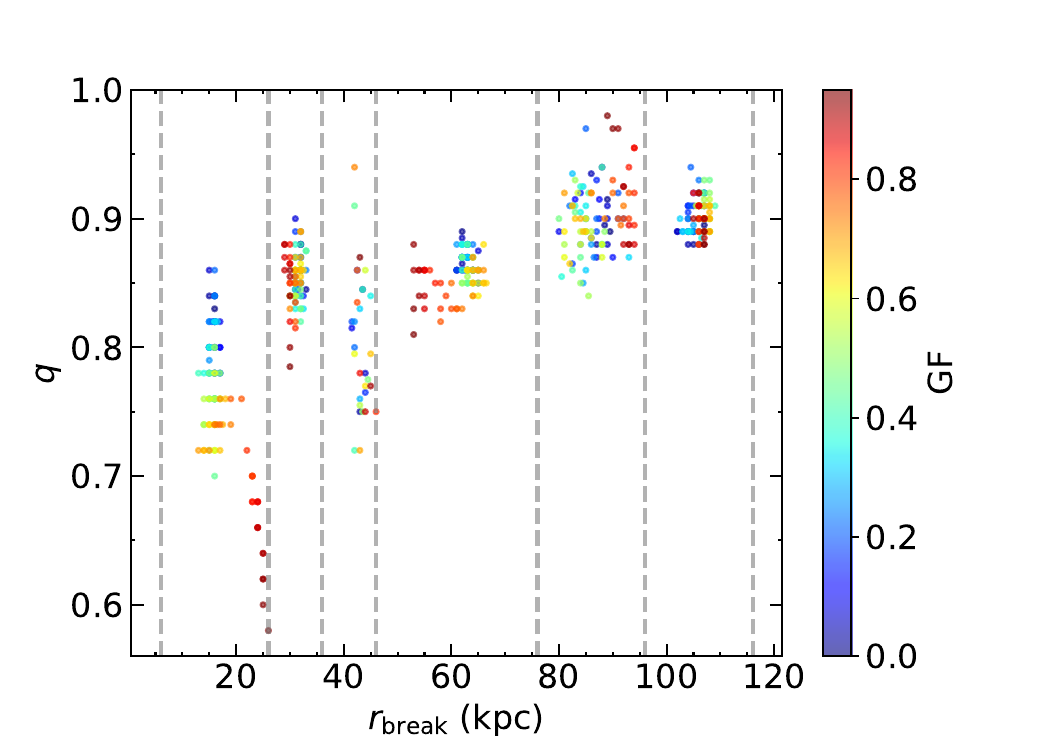}
    \caption{Goodness of fit as a function of $r_{\rm break}$ and $q$.
    The gray dashed lines represent the boundaries of all bins.}\label{rbreak_q}
\end{figure}

Since the density obtained by KDE is biased by the input parameters and the algorithm itself, the parameter values with $\rm GF_{max}$ of SBPL cannot fully represent the best-fitting results.
In order to get reliable break radii, the $\rm GF_{max}$ value of SPL is used as the goodness-of-fit threshold $\rm GF_{t}$.
Note that a fit with SPL in the $(36,46)\,{\rm kpc}$ bin is severely influenced by the overdensities from the Sgr stream (i.e., $\rm GF_{max}\,\textless\,0$), so $\rm GF_{t}$ is set to 0 in this bin.
Figure \ref{rbreak_GF} shows $r_{\rm break}$ as a function of goodness-of-fit, we notice that the error of $r_{\rm break}$ will become smaller beyond the $\rm GF_{max}$ values of SPL (dashed line).
Therefore, not only the $\rm GF_{max}$ value of SPL as $\rm GF_{t}$ could get a better fit than SPL, but also get the break radius with less uncertainty.
Figure \ref{rbreak_q} shows goodness-of-fit as a function of $ r_{\rm break}$ and $q$.
In the region of $6\,{\rm kpc}\,\textless\,r\,\textless\,26\,{\rm kpc}$, GF decreases with the increase of $q$, and the cut-off radius is distributed outside 10\,kpc, which shows that the inner halo is very flat, and the break at 10\,kpc does not exist.
In the 4D $(n,q,\delta n,r_{\rm break})$ space, we select the points with $\rm GF\,\textgreater\,GF_{t}$ to estimate the best-fitting values and 1$\sigma$ uncertainties using 16th, 50th and 84th percentiles, as listed in Table \ref{table_result},
which can provide a reference for the classical fitting method.

\subsection{Classical fitting method}\label{Classical fitting method}
This method is generally used in previous studies, and we found that this method can get the same conclusion as GFFM, which may exclude the influences of different fitting methods and several possibilities of breaks mentioned in Section \ref{Dataset and selection criteria}. 
We divide our clean sample into five bins to ensure that each bin contains possible break radii obtained by GFFM.
Furthermore, we use SPL, SBPL and the new doubly break power law (DBPL) to fit the data in order to further confirm that the determined breaks are real, and $(a,p,\gamma,\eta,\beta)$ are free parameters due to the acceptable computation time.
The normalized likelihood for the clean sample is as follows: \newline
\begin{equation}
\begin{aligned}
{\rm ln}\,\mathcal{L}\,(r,\theta,\phi\lvert\,\Theta) & = 
\sum_{i=1}^{n}{\rm ln}\,\mathcal{L}(r_{i},\theta_{i},\phi_{i}\lvert \Theta),
\end{aligned}
\end{equation}
\begin{equation}
{\rm ln}\,\mathcal{L}(r_{i},\theta_{i},\phi_{i}\lvert \Theta)
= {\rm ln}\,\frac{ \rho_{\rm halo}(r_{i},\theta_{i},\phi_{i}\lvert\,\Theta)S_{\rm all}(r_{i},\theta_{i},\phi_{i})\lvert \textbf{\textit{J}}_i \rvert}{\iiint_V \rho_{\rm halo}(r,\theta,\phi\lvert\,\Theta)S_{\rm all}(r,\theta,\phi)\lvert \textbf{\textit{J}} \rvert \,dr\,d\theta\,d\phi}, 
\end{equation}
\begin{equation}
S_{\rm{all}} = \prod_{j=1}^{k}S_{j}. 
\end{equation}
\newline
where $ S_{\rm{all}} $ represents the multiplication of all the above-mentioned selection functions in Section \ref{Dataset and selection criteria}.
The Jacobian term $\lvert \textbf{\textit{J}} \rvert=r^2{\rm sin}\,\theta$ reflects the transformation from $(x,\,y,\,z)$ to $(r,\,\theta,\,\phi)$ coordinates.
Note that $f=1$ in the selection functions of SMC, LMC, globular clusters, and the selection function of Sgr dwarf galaxy and stream follows equation (11).
Here, a prior is adopted on the basis of previous studies \citep[e.g.,][]{hernitschek2018profile,iorio2018first} as follows:
\newline
\begin{equation}
{\rm ln}\,p\,(\Theta) = \left\{
\begin{aligned}
0, & & {n\in(1,6),q\in(0,1),r_{\rm break}\in(r_{\rm down},r_{\rm up}),}\\
& & {\beta,\,\eta\,{\rm and}\,\lambda\in(-0.5\pi,0.5\pi),a\,{\rm and}\,p\in(1,+\infty)}\\
-\infty, & & {\rm{else.}}\\
\end{aligned}
\right.
\end{equation}\newline
where $ r_{\rm down}$ and $ r_{\rm up} $ denote the lower and upper boundaries of each bin, respectively. 
We perform all integrations in this work using the \textit{vegas} algorithm \citep{lepage1978new} through its Python implementation\footnote{\url{https://github.com/gplepage/vegas}}, in which $nitn$ is set to 10, and $neval$ is set to 1000. The final estimate of the integral is obtained from the average of $nitn$ \textit{vegas} runs with \textit{neval} integrand evaluations.
We sample the posterior probability over the parameter space with Goodman \& Weare's Affine Invariant Markov Chain Monte Carlo \citep[MCMC,][]{goodman2010ensemble} using the Python module \textbf{\textit{emcee}} \citep{foreman2013emcee}.
The final results are summarized in Table \ref{results from the classical fitting method}.

\subsection{Selection of the new merger candidates}\label{Metallicity correction and selection of the new merger}
In each bin, we use the new broken power law to fit the data in the two methods, and preliminarily determine the distribution of shell-like structures beyond $r\sim50\,{\rm kpc}$ according to the break radius in the density profile.
To explore high-purity candidates of shells from our clean sample, we cluster the objects in each bin (When assigning stars to bins in spherical $r$, the bin edges are selected so that each bin contains
$\rm N_{stars}=500$ objects because the clustering algorithm requires enough spatial information of stars as input.) twice by the HDBSCAN clustering algorithm \citep[Hierarchical Density-Based Spatial Clustering of Applications with Noise,][]{campello2013density}, in which we use the HDBSCAN Python package\footnote{\url{https://github.com/scikit-learn-contrib/hdbscan}} \citep{mcinnes2017hdbscan}.
The specific steps are as follows:
(\romannumeral1) we cluster the stars in the Galactic longitude and latitude $(l,b)$ space, and then remove stars marked with $-1$ and the clustering probabilities less than 0.6;
(\romannumeral2) in the heliocentric distance space, we cluster the stars retained from the step (\romannumeral1) in each bin, and then remove those stars marked with $-1$ and the clustering probabilities less than 0.8;
(\romannumeral3) among the remaining stars in the step (\romannumeral2), we select the stars located outside $r\sim50\,$kpc.

\section{Results}\label{Results}
Previous studies have shown that an accreted galaxy on a near-radial orbit, such as GSE, will result in ring-like or shell-like structures, corresponding to apocenter pile-ups in the density map or breaks in the density profile.
They are, respectively, located at the penultimate and final apocentric distances before it fully merged with the MW as well as the edges of the ultra-dense cores in the substructures after merging.
Therefore, we may use the breaks in the broken power law to find shell-shaped structures which are remnants of an accreted galaxy on a near-radial orbit.

\subsection{Analysis of results by GFFM}

\begin{figure}
	\includegraphics[width=\columnwidth]{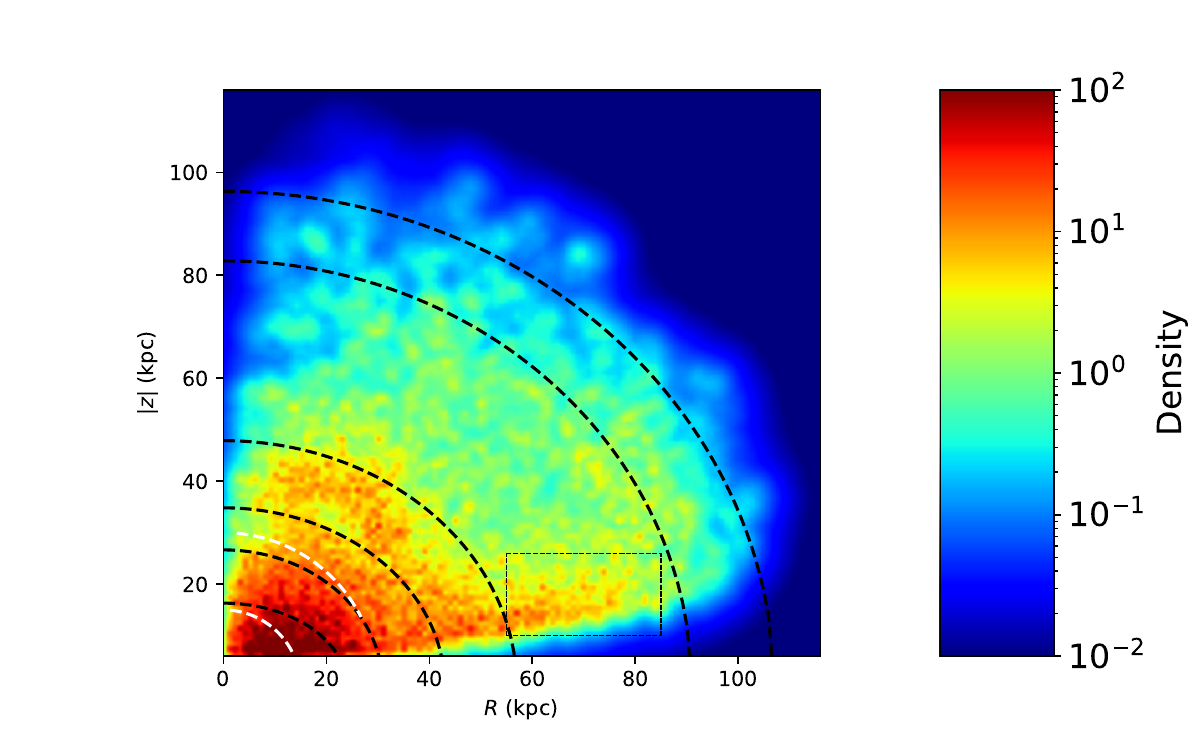}
    \caption{
    Cylindrical map showing all break radii and the distribution of the number density of the total sample.
    The black dashed lines represent all break radii explored by GFFM, 
    and the white dashed lines represent the inner and outer boundaries of the GSE apocenters \citep{malhan2022global}.}\label{allbreaks}
\end{figure}

\begin{table}
	\centering
	\caption{The best-fitting parameters obtained by GFFM.}
	\label{table_result}
	\begin{tabular}{lcccccc}
		\hline
		range (kpc) & $r_{\mathrm{break}}\,({\rm kpc})$ & $n$ & $q$ & $\delta n$ & $\rm GF_{t}$\\
		\hline
		(6,26) & $24_{-3}^{+2}$ & $2.1_{-0.3}^{+0.3}$ & $0.68_{-0.10}^{+0.08}$ & $0.5_{-0.2}^{+0.3}$ & 0.811 \\
        (26,36) & $31_{-3}^{+3}$ & $2.9_{-0.2}^{+0.3}$ & $0.86_{-0.11}^{+0.10}$ & $0.5_{-0.2}^{+0.3}$ & 0.628 \\
        (36,46) & $43_{-2}^{+2}$ & $3.4_{-0.1}^{+0.1}$ & $0.81_{-0.09}^{+0.11}$ & $-0.3_{-0.1}^{+0.1}$ & 0 \\
        (46,76) & $57_{-7}^{+10}$ & $3.2_{-0.1}^{+0.1}$ & $0.84_{-0.10}^{+0.11}$ & $0.2_{-0.1}^{+0.2}$ & 0.746 \\
        (76,96) & $91_{-7}^{+4}$ & $3.4_{-0.1}^{+0.1}$ & $0.91_{-0.07}^{+0.07}$ & $0.2_{-0.1}^{+0.1}$ & 0.673 \\
        (96,116) & $107_{-7}^{+6}$ & $3.8_{-0.2}^{+0.1}$ & $0.90_{-0.07}^{+0.07}$ & $0.4_{-0.2}^{+0.2}$ & 0.281 \\
		\hline
	\end{tabular}
\end{table}

The six breaks explored by GFFM are shown in the top panel of Figure \ref{allbreaks}. 
We consider that the inner and outer boundaries of the apocentric distance for the GSE merger in \citet{malhan2022global} are equivalent to the GSE penultimate and final apocentric distances in \citet{naidu2021reconstructing} for the following reasons:
\cite{malhan2022global} used globular clusters, dwarf galaxies, and stellar streams that well preserve the dynamical properties of its progenitor galaxy as their sample; 
\citet{naidu2021reconstructing} predicted that two breaks occur at the penultimate and final apocenters of the GSE, which are consistent with the upper and lower boundaries found by \citet{malhan2022global};
\citet{han2022stellar} recently confirmed that the predictions of \citet{naidu2021reconstructing} by fitting a tilted triaxial ellipsoid with a doubly broken power law along its flattened radius and performing N-body simulations;
the apocentric distance should become closer to the Galactic centre over time before it is completely accreted. 
As shown in Figure \ref{allbreaks}, the two apocenters of high-eccentricity GSE merger (white dashed line) closely match the two breaks at $r_{\rm e}=24$ and 31\,kpc.
It is clear that a reverse break at $r_{\rm e}\sim43\,{\rm kpc}$ with negative $\delta n$ is caused by the Sgr stream at large zenithal angles.
A structure composed of several overdensities at low zenithal angles is shown in the black box in Figure \ref{allbreaks}.
After removing the Sgr stream, the intersecting area between the inner over-density area and the outer structure at low zenithal angles ($40\,{\rm kpc}\,\textless\,R\,\textless\,60\,{\rm kpc}$ and $\lvert z\rvert\,\textless\,20\,{\rm kpc}$) is still visible, as shown in the bottom panel of Figure \ref{figR_zall}, so the structure may lead to the break at $r_{\rm e}\sim57\,{\rm kpc}$ in the density profile.
However, in Figure \ref{allbreaks}, it can be found that the break at $r_{\rm e}\sim57\,{\rm kpc}$ is just at the outer boundary of the over-density region in the Sgr stream, so the break could also be attributed to the Sgr stream.
The break at $r_{\rm e}\sim91\,{\rm kpc}$ could also be derived by the outer boundary of the structure at low zenithal angles.
From Figure \ref{allbreaks} we can see that the density beyond the farthest break ($r_{\rm e}\sim107\,{\rm kpc}$) is extremely sparse, which is caused by the observation limit.
In addition, from Table \ref{table_result} It can be seen that the inner halo is very flat (small value of $q$).
\subsection{Analysis of results by the classical fitting method}\label{Analysis of results by the classical fitting method}

\begin{figure*}
	\includegraphics[width=0.66\textwidth]{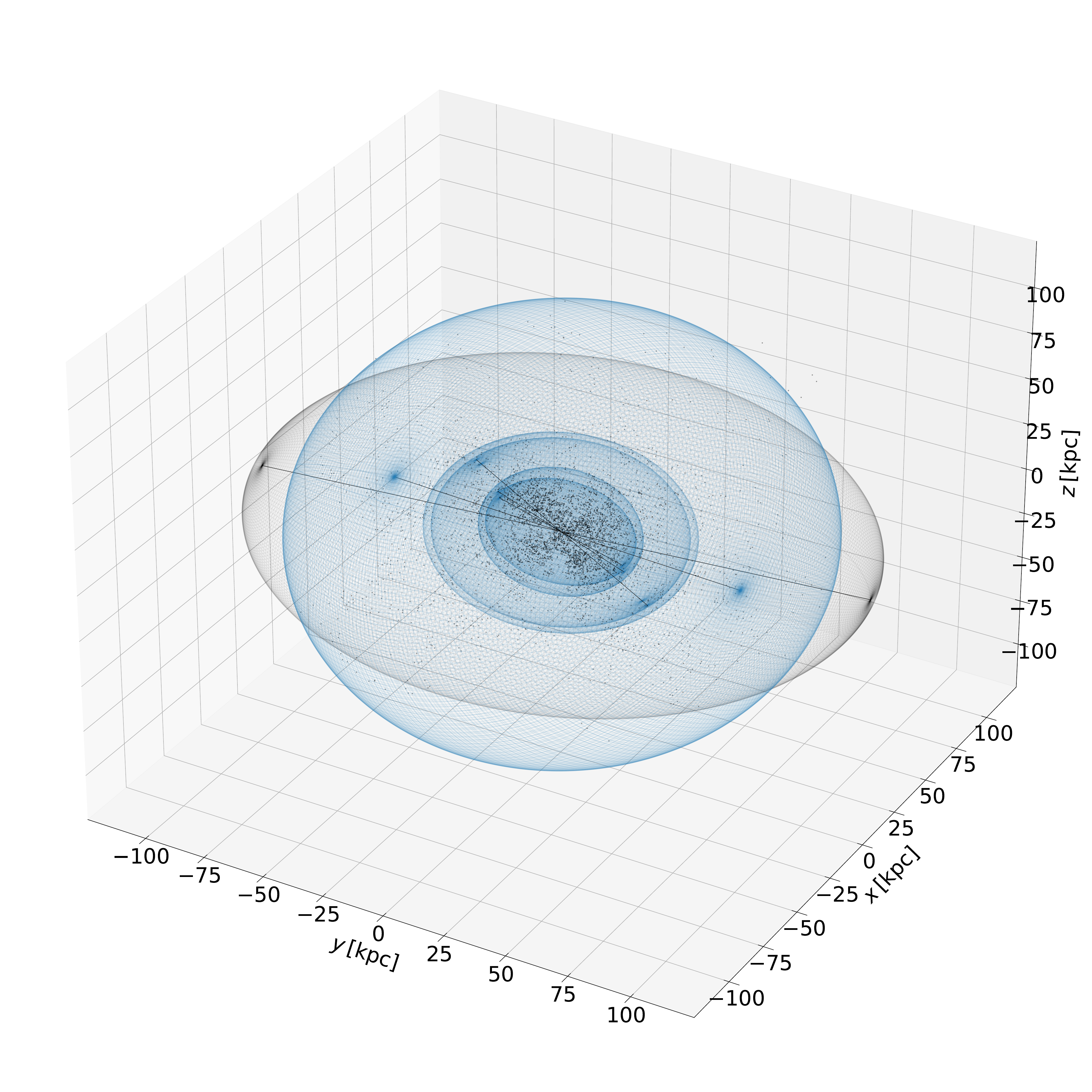}
    \caption{All break radii explored by the classical fitting method, including $r_{\rm break}\sim 22.99$\, kpc, 27.18\, kpc, 45.25\, kpc, 48.05\, kpc, 92.99\,kpc and 103.77\,kpc, are represented by triaxial ellipsoids with major axes indicated by black solid lines. 
    We plot the ellipsoid corresponding to the break at $92.99\,{\rm kpc}$ in black, which has a clearly distinct major axis and flattening at a larger distance.
    The black dots represent all overdensities explored by the HDBSCAN clustering algorithm.
    }\label{rbreak_group}
\end{figure*}

It is found that the flatness in the outer halo is extremely affected by overdensities due to the low-density background halo, so we also fit the clean sample.
In order to further confirm that the structure does not belong to the Sgr stream from 20 to over 100\,kpc \citep{belokurov2014precession,sesar2017100,hernitschek2017geometry,ramos2020full}, and to rule out the influences of remnants of LMC, SMC and globular clusters, 
we apply the classical fitting method for the clean sample and study the variation of various parameters with $r$ to confirm the spatial extent occupied by the structure at low zenithal angles.
In Table \ref{results from the classical fitting method} we summarize the results obtained by fitting the clean sample with the SPL, SBPL and DBPL density profiles including the additional five free parameters ($a,p,\gamma,\eta,\beta$).
In Figure \ref{rbreak_group}, we show the members of all substructures obtained by the HDBSCAN clustering algorithm and all triaxial ellipsoids, namely all breaks.
We can find that the direction of the major axis in the region of $r\,\textless\,66\,{\rm kpc}$ seems to be consistent with the elongation direction of the diffuse substructure in the inner halo ($\textless\,36\,{\rm kpc}$), which is in agreement with previous studies.
\citet{iorio2019shape} have found that the two large diffuse overdensities, namely HAC and VOD, are aligned with the semi-major axis of the halo within $\sim30\,{\rm kpc}$ from the Galactic centre.
The major axis of the triaxial ellipsoid is nearly coincident with the final two apocenters of the GSE in N-body simulation \citep{naidu2021reconstructing}, indicating that the major axis is dominated by the penultimate and final apocenters of the GSE merger.

Here we evaluate the validity and reliability of breaks in each interval and compare the results of different models to determine which of them gives the best description of the data by the Bayesian evidence.
Assuming that the posterior distributions are approximately Gaussian, the Bayesian evidence can be estimated by the Bayesian Information Criterion \citep[BIC,][]{Schwarz1978} defined as
\begin{equation}
{\rm BIC}=-2{\rm ln}(\mathcal{L}_{\rm max})+k{\rm ln}N_{\rm s}
\end{equation}
where $k$ is the number of free parameters, $N_{\rm s}$ is the data sample size and $\mathcal{L}_{\rm max}$ is the maximum value of the likelihood.
The BIC is commonly utilized for comparing models of varying parameter dimensions, with preference given to the model exhibiting the lowest BIC value.
In Table \ref{results from the classical fitting method}, we notice that the BIC values for SBPL in $(6,26)\,{\rm kpc}$, $(26,36)\,{\rm kpc}$ and $(66,96)\,{\rm kpc}$ are smaller than those for SPL, which indicates that the SBPL model is more suitable for the data than SPL in these intervals, resulting in increased confidence about the presence of breaks within these intervals.

\begin{figure}
	\includegraphics[width=\columnwidth]{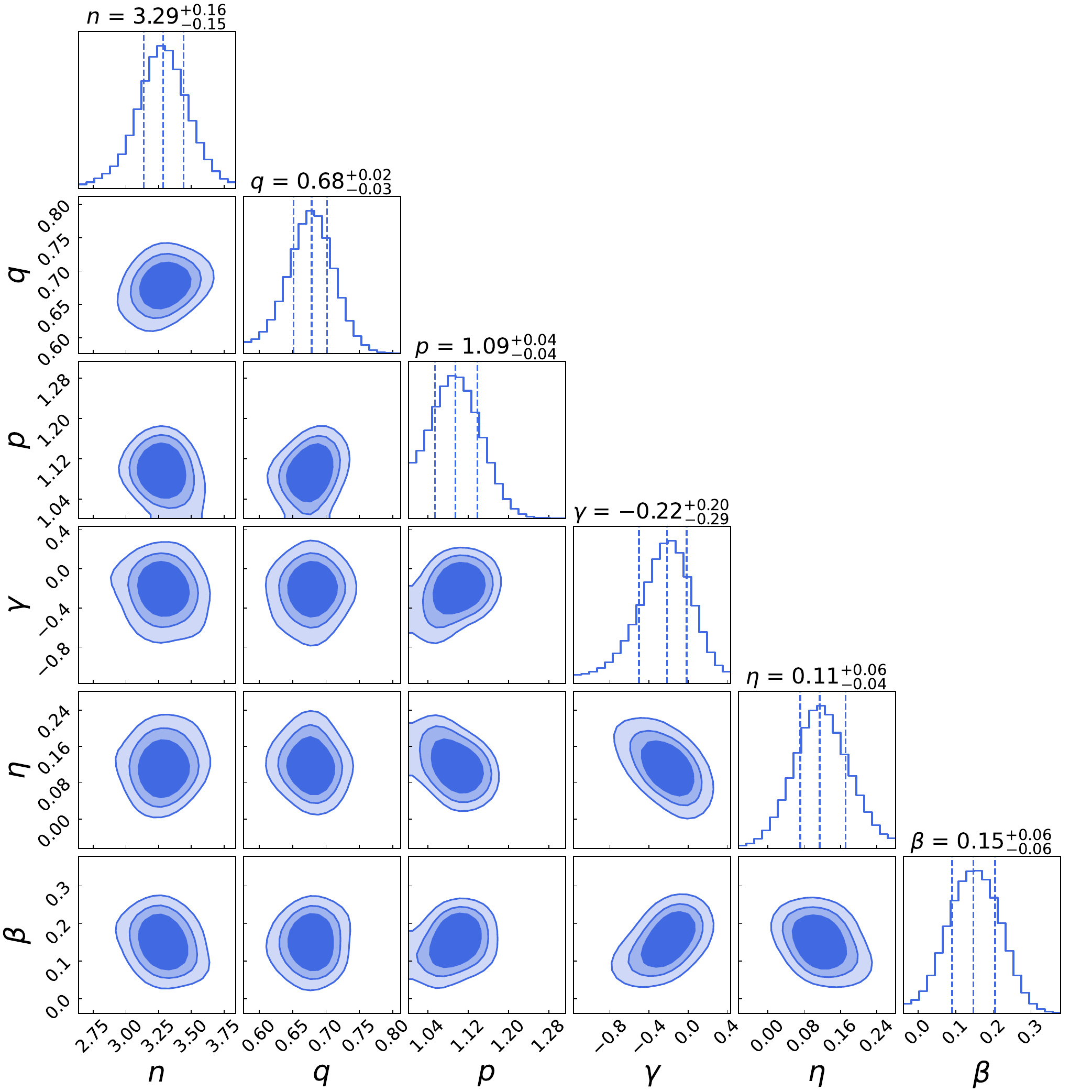}
    \caption{The probability distribution shown in the figure is obtained by fitting the SPL model to the clean sample within $36\,{\rm kpc}\,\textless\,r\,\textless\,66\,{\rm kpc}$ by our application of the classical fitting method.}\label{33_66_0break}
\end{figure}

\begin{figure*}
	\includegraphics[width=\textwidth]{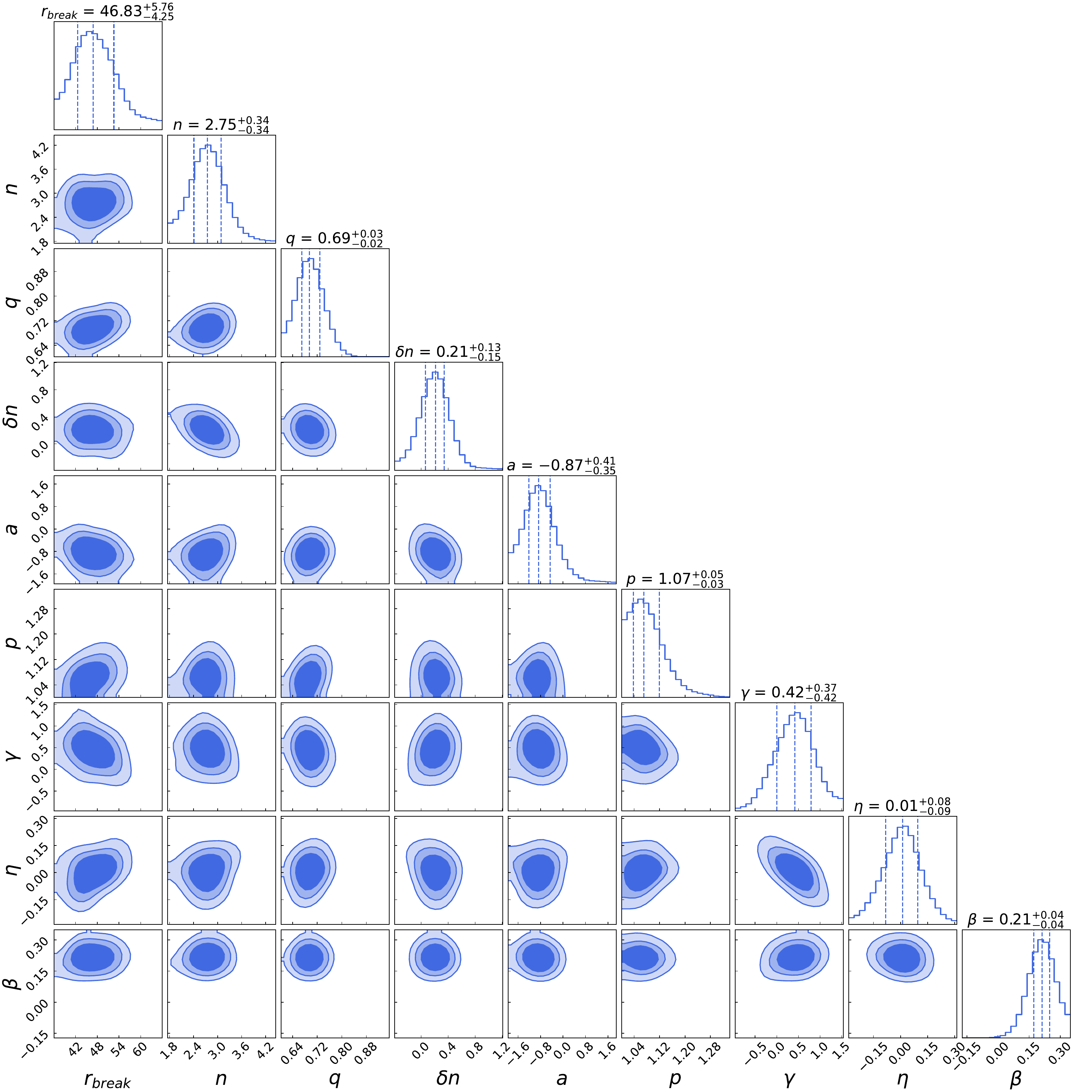}
    \caption{Similar to Figure \ref{33_66_0break}, but using the SBPL model here.}\label{33_66_1break}
\end{figure*}

\begin{figure*}
	\includegraphics[width=\textwidth]{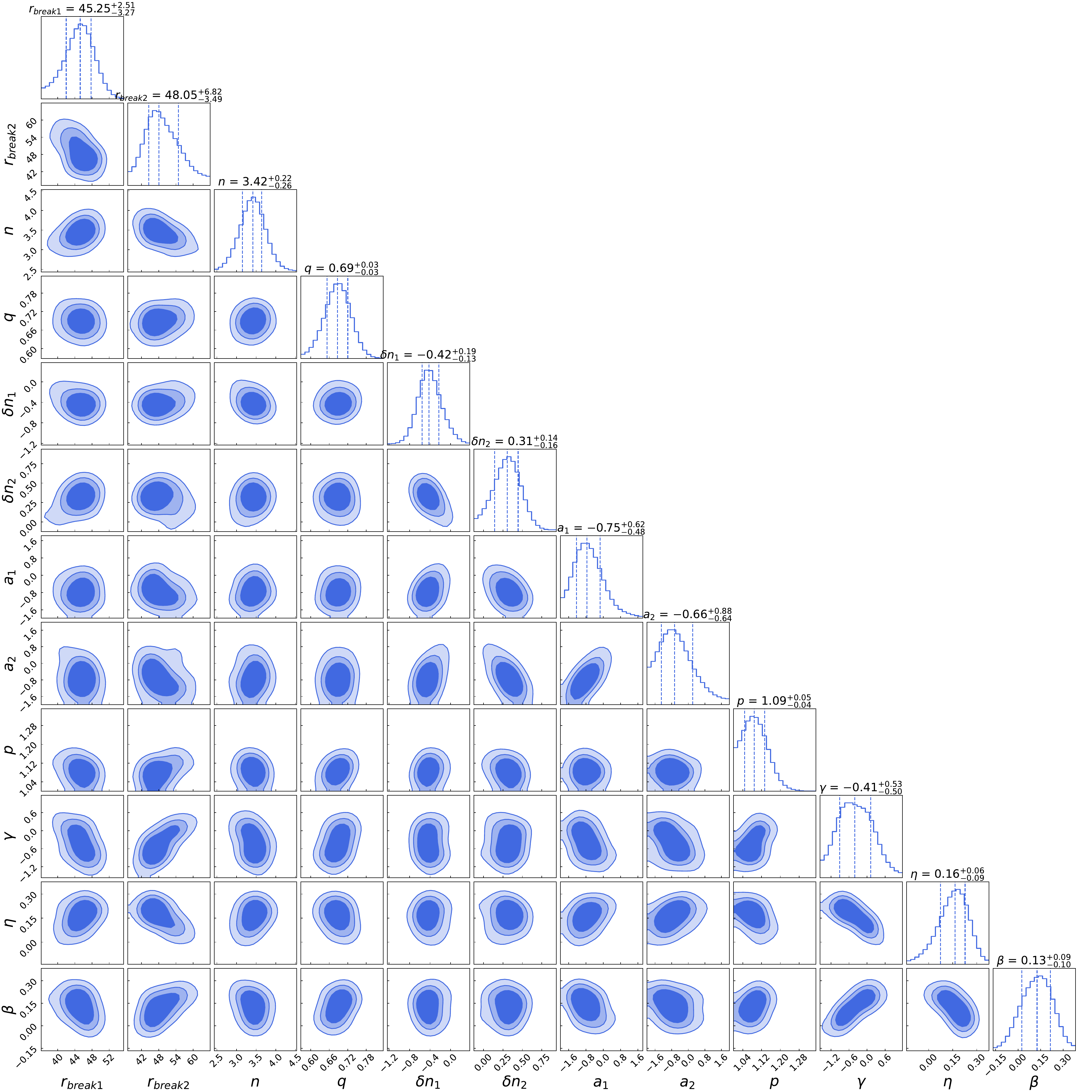}
    \caption{Similar to Figure \ref{33_66_0break}, but using the DBPL model here.}\label{33_66_2break}
\end{figure*}

It is known that the break at $r\sim20\,{\rm kpc}$ confirmed by \citet{deason2018apocenter} is caused by the GSE apocenter pile-up, however, some previous studies \citep{xue2015radial,iorio2018first} have shown that the BIC value obtained for BPL is larger than that for SPL.
The BIC is similar to the maximum-likelihood criterion, but it takes into account a penalty depending on the number of free parameters $k$, such that for two models with the same likelihood, the one with more parameters is penalized. 
In other words, a model with more parameters will be unnecessary if two models, with distinct numbers of free parameters, comparably fit the data well.
Therefore, we cannot deny the presence of breaks just based on larger BIC.
In the region of $36\,{\rm kpc}\,\textless\,r\,\textless\,66\,{\rm kpc}$, the BIC of the SPL model is smaller than those obtained for SBPL and DBPL due to large break scales, for example, the break scales with $a=-0.87^{+0.41}_{-0.35}$ in SBPL and $a_1=-0.75^{+0.62}_{-0.48},\ a_2=-0.66^{+0.88}_{-0.64}$ in DBPL ($a\sim-0.87,\ -0.75$ and $-0.66$ mean that $n$ can only increase by $0.4\delta n$ within $r_{\rm break}\pm5.39,4.09$ and $3.32\,{\rm kpc}$, respectively.) are much larger than all breaks except the farthest break caused by the observation limit.
In Figures \ref{33_66_0break}-\ref{33_66_2break}, it can be found that the major axis direction in SBPL, namely $(\gamma,\eta)=(0.42^{+0.37}_{-0.42},0.01^{+0.08}_{-0.09})$, is obviously different from those of SPL $(-0.22^{+0.20}_{-0.29},0.11^{+0.06}_{-0.04})$ and DBPL $(-0.41^{+0.53}_{-0.50},0.16^{+0.06}_{-0.09})$.
In addition, the distribution of $\gamma$ in DBPL is quite wide, ranging from $-0.91$ to 0.12.
Furthermore, we find that the density distribution in the region of $36\,{\rm kpc}\,\textless\,r\,\textless\,66\,{\rm kpc}$ is significantly different from that in the inner halo $(r\,\textless\,36\,{\rm kpc})$, that is, its flatness suddenly increases in this interval, which is consistent with the results of the three models $(q\sim0.69)$.
Combined with the above-mentioned results, it shows that the breaks at $r_{\rm e}\sim45.25\,{\rm kpc}$ and $48.05\,{\rm kpc}$ are caused by two diffuse substructures distributed in different directions, namely $(\gamma\,\textgreater\,0,\ \eta\sim0\ {\rm and}\ \gamma\,\textless\,0,\eta\sim0.1)$, one of which is the debris of GSE due to the tendency towards the major axis direction in the inner halo ($r\,\textless\,36\,{\rm kpc}$), we also plot these two breaks in Figure \ref{rbreak_group}.


Here we analyze the fitting results in $66\,{\rm kpc}\,\textless\,r\,\textless\,96\,{\rm kpc}$.
In Figure \ref{rbreak_group}, we notice that the major axis with small uncertainties in $66\,{\rm kpc}\,\textless\,r\,\textless\,96\,{\rm kpc}$ $(\gamma\,\textgreater\,0,\ \eta\sim0)$ is obviously different from that caused by GSE in the inner halo $(\gamma\,\textless\,0,\ \eta\sim0.1)$.
Therefore, the diffuse overdensity corresponding to this major axis $(\gamma\,\textgreater\,0,\ \eta\sim0)$ is dominant within $66\,{\rm kpc}\,\textless\,r\,\textless\,96\,{\rm kpc}$.
It could be the over-density structure in the black box, as shown in Figure \ref{figR_zall}. Since its distribution at low zenithal angles makes larger flatness in the region of $66\,{\rm kpc}\,\textless\,r\,\textless\,96\,{\rm kpc}$ than that in the inner halo $(r\,\textless\,36\,{\rm kpc})$, namely smaller $q$, as shown by the black ellipsoid in Figure \ref{rbreak_group}.
In the region of $36\,{\rm kpc}\,\textless\,r\,\textless\,66\,{\rm kpc}$, the flattening of $\sim0.69$ resulted from the fact that the number density is extremely high at low zenithal angles but is very sparse in the region without the Sgr stream at large zenithal angles, as shown in the bottom panel of Figure \ref{figR_zall}.
The over-density structure distributed at lower zenithal angles could be the `eccentric Cloudy' suggested by \citet{johnston2008tracing}, which occupies a larger range with distances of $20-100\,{\rm kpc}$.
In Figure \ref{rbreak_group}, we can see that, within 96\,kpc$\,\textless\,r\,\textless\,116$\,kpc, the value of $q$ reaches a reasonable level based on the relationship between $q$ and $r$ in the inner halo $(r\,\textless\,36\,{\rm kpc})$, which indicates that there are a few members of the over-density structure at low zenithal angles in this region.

We now analyze all the breaks in Figure \ref{rbreak_group}:
(\romannumeral1) The farthest break at $r_{\rm e}\sim104\,$kpc could be caused by the observation limit due to its largest scale ($a\sim-1.02$ means that $n$ just increases by $0.4\delta n$, i.e. $0.4\times1.54$, within $104\pm7.61\,{\rm kpc}$).
(\romannumeral2) The break at $r_{\rm e}\sim93\,$kpc ($a\sim0.44$ means that $n$ increases by $0.8\delta n$, i.e. $0.8\times(-0.22)$, just within $93\pm1.12\,{\rm kpc}$) could correspond to some compact substructures due to its smallest scale.
(\romannumeral3) For the best-fitting results of DBPL in $36\,{\rm kpc}\,\textless\,r\,\textless\,66\,{\rm kpc}$, the reverse break at $r_{\rm e}\sim45\,$kpc could be created by debris of GSE, and the other break at $r_{\rm e}\sim48\,$kpc could be caused by some diffuse substructures at low zenithal angles.
The best-fitting results for SBPL show that the major axis (i.e., $\gamma\sim0.42,\ \eta\sim0.01$) obviously deviates from that dominated by HAC and VOD in the inner halo $(\gamma\,\textless\,0,\ \eta\sim0.1)$, and that the break at $r_{\rm e}\sim47\,{\rm kpc}$ with $\delta n \sim0.21$ is very similar to the break at $r_{\rm e}\sim48\,{\rm kpc}$ with $\delta n\sim0.31$ in DBPL, indicating that unknown substructures at low zenithal angles could lead to the break around $48\,{\rm kpc}$ in the density profile and the distinct major axis in SBPL.
Based on the two breaks at $r_{\rm e}\sim48\,{\rm kpc}$ and $\sim93\,{\rm kpc}$, we are able to preliminarily locate these unknown overdensities at low zenithal angles, which contributes to the flattened halo and the orientation of the major axis approaching $(\gamma\,\textgreater\,0,\ \eta\sim0)$.
However, the best-fitting results for SPL show that the major axis is consistent with that dominated by HAC and VOD, indicating that there is debris from them in this range.
In addition, the best-fitting results for DBPL show that the major axis is roughly consistent with that dominated by HAC and VOD, while the uncertainty is significant (i.e., $\gamma=-0.41^{+0.53}_{-0.50}$).
This could be caused by the combined effects of unknown substructures and fragments from HAC and VOD.

Finally, we compare the results of the two methods. 
In the GFFM approach, the setting of $f$ in the selection function of LMC, SMC, the Sgr dwarf galaxy and globular clusters is equivalent to including those substructures in the corresponding space regions, and GFFM is used for the total sample, so the slope $n$ in the fitting results of GFFM is lower than that of the classical fitting method.
In addition, the fitting results of the total sample will inevitably get a larger $z\text{-to-}x$ ellipsoid axial ratio $q$, because of a clear decrease of density at large zenithal angles after removing the Sgr stream, as shown in Figure \ref{figR_zall}.
Furthermore, if we fit the total sample with the classical fitting method, a smaller $y\text{-to-}x$ ellipsoid axial ratio $p$ should be derived in the region of $D_{\rm sun}\textgreater15\,$kpc, since its stellar density distribution elongates along the $x$-axis on the $x\text{-}y$ plane or flattens along the $y$-axis. 

\subsection{The metallicity distribution as a function of radius}\label{analysis of metallicity distribution function}

\begin{figure}
	\includegraphics[width=\columnwidth]{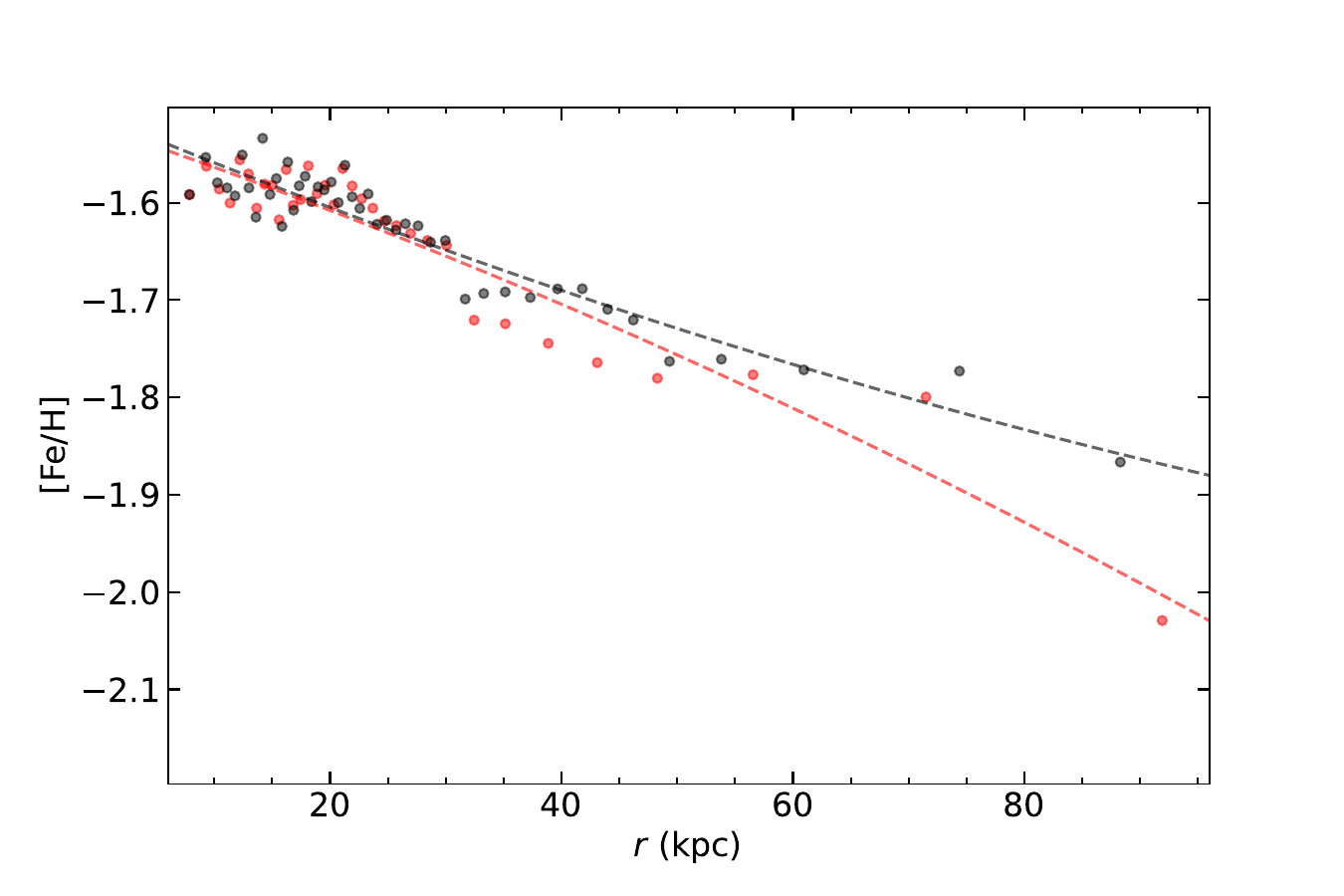}
    \includegraphics[width=\columnwidth]{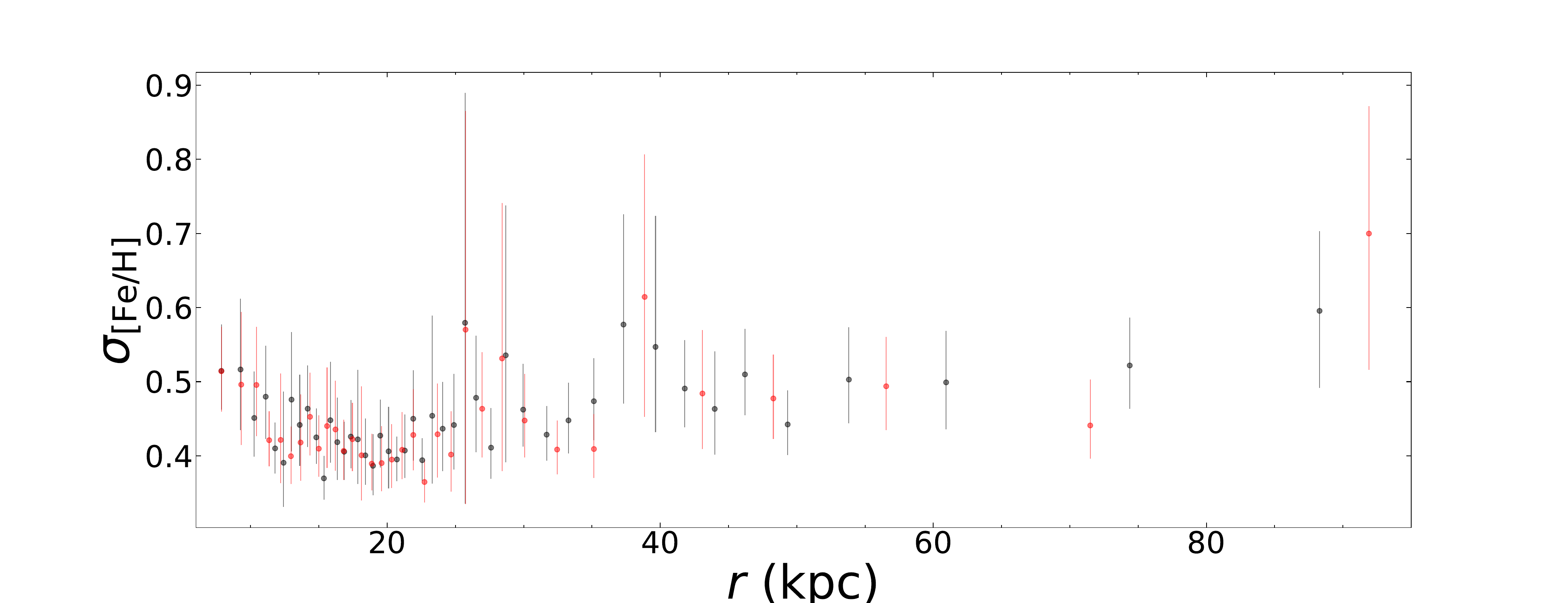}
    \caption{Matellicity (top) and its dispersion (bottom) as functions of spherical radius $r$.
    The black (red) dots represent the metallicity derived from $P$ and $\Phi_{31}$ of the total (clean) sample. 
    The black (red) dashed line denotes the second-order polynomial fitted to the black (red) dots using the least square method.
    We assign stars with the \textit{Gaia} $\Phi_{31}$ to bins in spherical $r$ so that each bin contains $\rm N_{stars}=500$ objects. 
    We get the statistical errors by computing $95\%$ confidence intervals of the 10000 bootstrapped resamples of the data in each bin.}\label{figr_feh_c}
\end{figure}

\begin{figure}
    \includegraphics[width=\columnwidth]{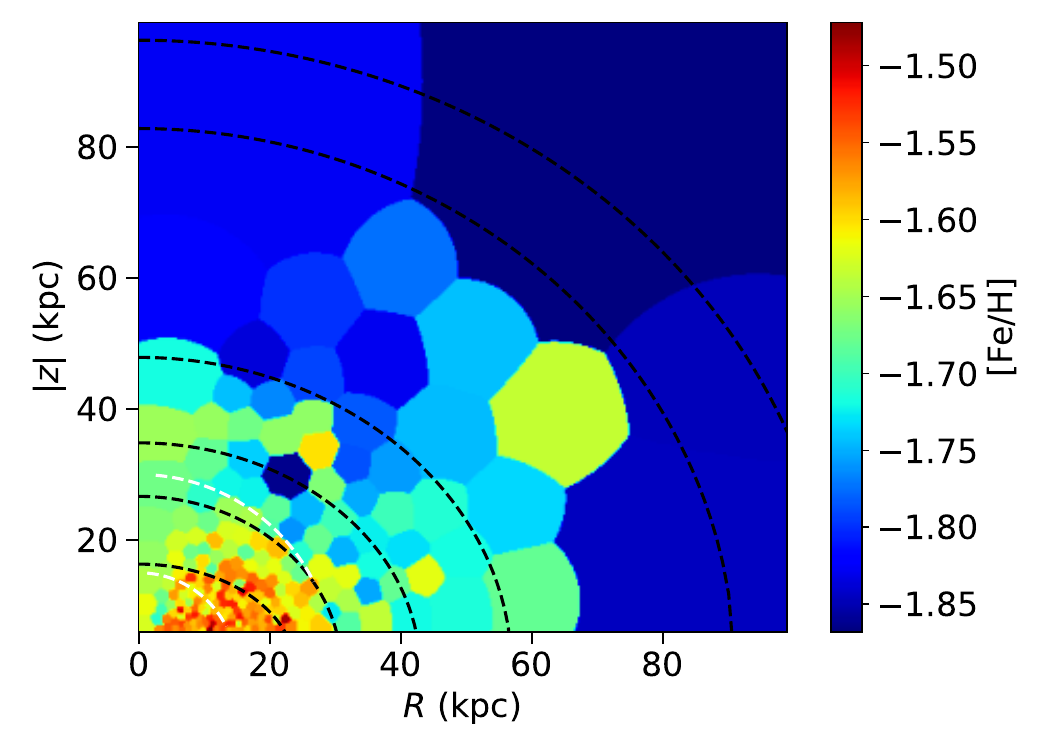}
	\includegraphics[width=\columnwidth]{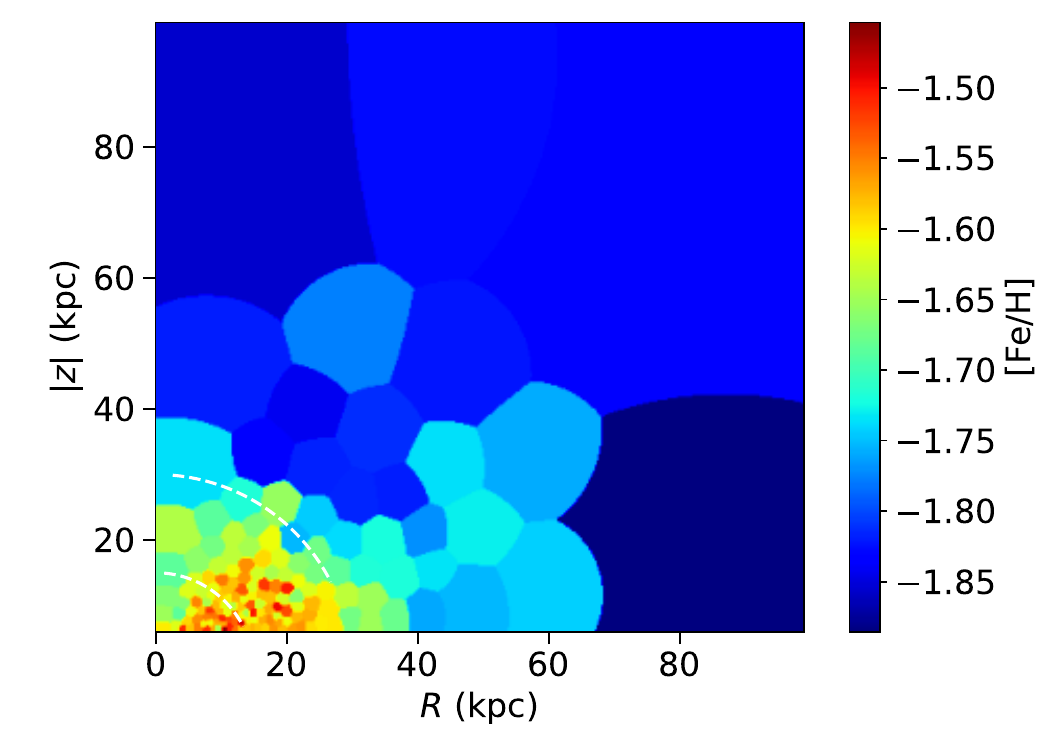}
    \caption{Cylindrical maps showing the distributions of metallicities of stars with $\Phi_{31}$ in the total sample (top) and clean sample (bottom).
    The black dashed lines represent all break radii explored by GFFM, including $r_{\rm break}\sim24\,{\rm kpc}$, 31\,kpc, 43\,kpc, 57\, kpc, 91\,kpc and 107\,kpc.
    The white dashed lines represent the inner and outer boundaries of the GSE apocenters \citep{malhan2022global}.}\label{vorbinfeh}
\end{figure}

\begin{figure*}
\centering
	\includegraphics[width=\textwidth]{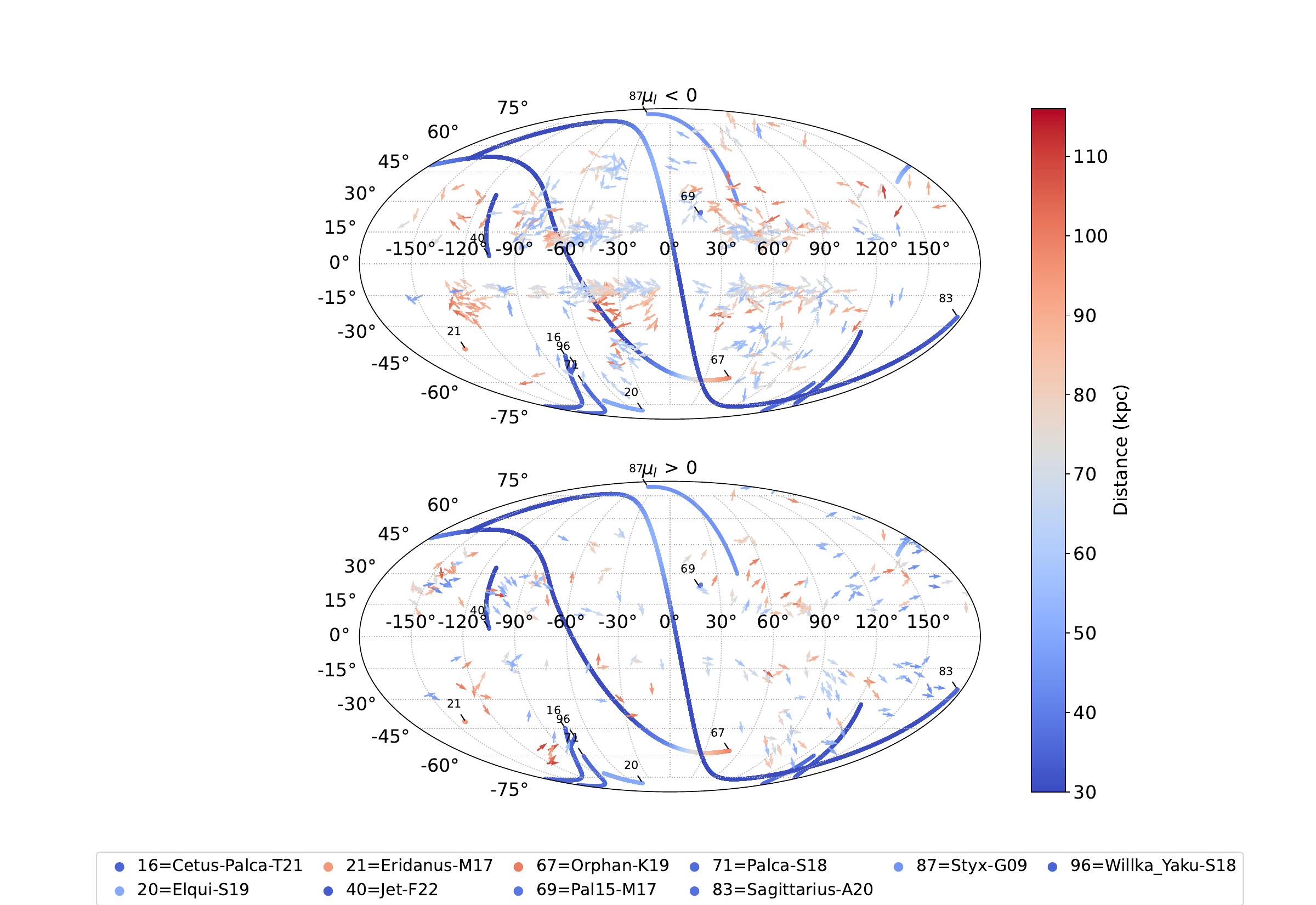}
    \caption{The distribution of the candidates of the shell-shaped structure and distant streams (heliocentric distances greater than 30\,kpc) in Galactic coordinates. The arrows indicate the direction of the proper motion and are color-coded according to the stars' distance.}\label{merger_pmlb}
\end{figure*}
\begin{figure*}
	\includegraphics[width=\textwidth]{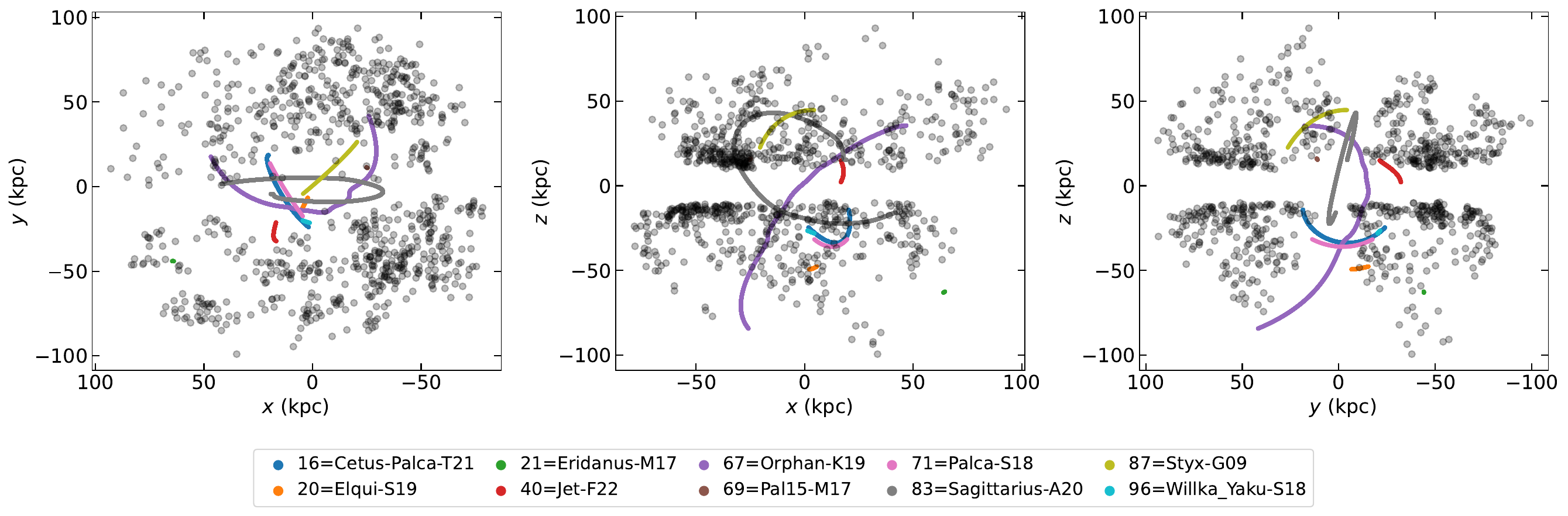}
    \caption{Spatial locations of the candidates of the shell-shaped structure plotted in Figure \ref{merger_pmlb} and distant streams in Cartesian coordinates.}\label{mergerxyz}
\end{figure*}
\begin{figure*}
    \includegraphics[width=\textwidth]{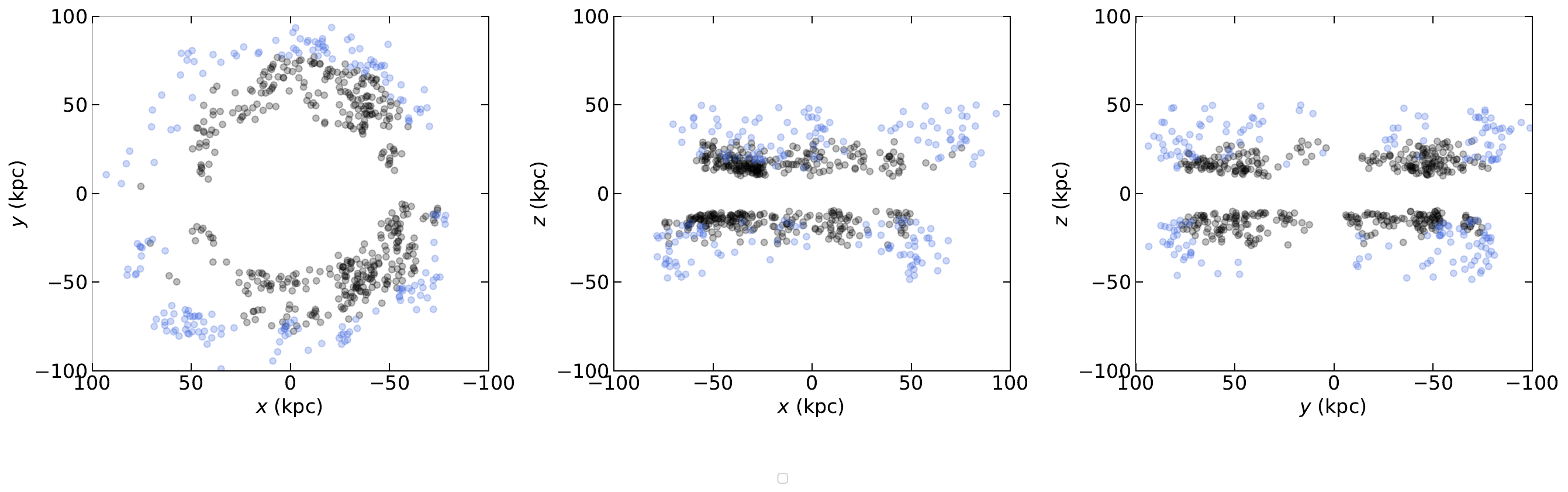}
    \caption{In order to show the shell-shaped structure at low zenithal angles more clearly, we further select them to be members in the outer halo in Figures \ref{merger_pmlb} and \ref{mergerxyz}. 
    The black dots represent members satisfying $50\,{\rm kpc}\,\textless\,r\,\textless\,80\,{\rm kpc}$ and $|z|\,\textless\,30\,{\rm kpc}$ in Figure \ref{mergerxyz}.
    The royalblue dots represent members satisfying $r\,\textgreater\,80\,{\rm kpc}$ and $|z|\,\textless\,50\,{\rm kpc}$ in Figure \ref{mergerxyz}.}\label{shells_xyz}
\end{figure*}

It is important to explore the correlation of metallicity with breaks.
The metallicity derived from $P$ and $\Phi_{31}$ as a function of $r$, as shown in the top panel of Figure \ref{figr_feh_c}, can also imply the existence of some substructures.
Since the majority of stars in the region of $r\,\textless\,20\,{\rm kpc}$ are composed of the metal-rich GSE, high $\alpha$-disk and in-situ halo \citep{naidu2020evidence}, the median metallicity is about $\rm [Fe/H]\sim-1.58\,dex$ out to 20\,kpc.
In the region of $20\,{\rm kpc}\,\textless\,r\,\textless\,30\,{\rm kpc}$, there are many metal-poor stars belonging to LMS-1/Wukong, Sequoia, I’itoi and other small mergers, while the fraction of metal-rich stars belonging to high $\alpha$-disk and in-situ halo and GSE has decreased in this range \citep{naidu2020evidence}, which could together determine a gentle metallicity gradient.
The apocenter of the metal-rich Sgr leading tail \citep[$r_{\rm apo}=47.8\pm0.5$\,kpc,][]{belokurov2014precession} lies between 30\,kpc and 50\,kpc, and its members could be more dispersive due to their lower eccentricities.
Therefore, for the clean sample without the Sgr stream (red), we can clearly see the lower metallicity than that of the total sample (black) in this range, as shown in the top panel of Figure \ref{figr_feh_c}.
We attribute the turning points of $r\sim50\,$kpc and 80\,kpc to the over-density structure at low zenithal angles originating from the apocenter pile-ups of its progenitor galaxy.
It can be inferred that there may be one apocenter pile-up at each turning point of the metallicity, such as $r\sim20\,{\rm kpc}$, 30\,kpc (GSE), 40\,kpc (the Sgr leading stream), 50\,kpc and 80\,kpc (the over-density structure at low zenithal angles).
In other words, each turning point can imply a potential alteration of the fraction of substructure due to its apocenter pile-up, but this needs to be further validated using the complete data with 6D phase-space measurements and chemical information.

Here, we analyze the metallicity dispersion as a function of $r$ shown in the bottom panel of Figure \ref{figr_feh_c}.
The metallicity dispersion reaches a minimum around both $r\sim20$\,kpc and $\sim32\,{\rm kpc}$, which also indicates that the stars here are dominated by one component, namely the apocenters pile-up of GSE.
However, the mixture of multiple components with very distinct metallicities in approximately the same proportions will lead to a large dispersion.
Three predominant components in the region of $6\,{\rm kpc}\,\textless\,r\,\textless\,20$\,kpc, including in-situ halo, high-$\rm \alpha$ disk and GSE, lead to a gentle drop in metallicity dispersion, which can reflect the increase in the relative fraction of GSE and the decrease in the relative fraction of in-situ halo and high-$\rm \alpha$ disk.
In the region of $20\,{\rm kpc}\,\textless\,r\,\textless\,30\,{\rm kpc}$,
it leads to a prominent dispersion that the metal-poor background halo and a fraction of the metal-rich GSE less than that in the region of $r\approx15-20\,{\rm kpc}$.
For both total (black) and clean (red) samples, the peak around $r\sim40\,$kpc and the rise around 80\,kpc could be attributed to some unknown structures or selection effects.

In order to visualize the metallicities of some structures, such as the Sgr stream, the over-density structure at low zenithal angles, and the metallicity distributions around breaks, we show the metallicity distribution of stars from the total sample (top) and clean sample (bottom) in the $R-|z|$ space in Figure \ref{vorbinfeh}.
We separate the total sample and clean sample into many cylindrical $R$, $|z|$ bins with an average Poisson signal-to-noise ratio of 10 using the \textit{\textbf{vorbin}} python package \citep{vorbin2003}.
The top panel of Figure \ref{vorbinfeh} shows that the break at 24\,kpc is close to the edge of the metal-rich area $(r\,\textless\,20\,{\rm kpc})$, and that seven metal-rich bins are located around the break at 31\,kpc. These results could be the imprints of two apocenter pile-ups of GSE \citep{naidu2021reconstructing} and lead to a gentle metallicity gradient in $20\,{\rm kpc}\,\textless\,r\,\textless\,30\,{\rm kpc}$.
In Figure \ref{vorbinfeh}, the Sgr stream at large zenithal angles between two breaks at $r_{\rm e}\sim43\,$kpc and 57\,kpc is clearly metal-rich, and after it is removed, a large number of metal-poor background halo stars with $\rm [Fe/H]\sim-1.85$ dex is exposed.
There are members of minor mergers that result in faster size growth, and they contribute little mass to the Galactic centre but a lot to the outer halo \citep{karademir2019outer}.
In the bottom panel of Figure \ref{vorbinfeh}, it can be seen that the metallicity in the corresponding region of the over-density structure at low zenithal angles is significantly higher than that of the background halo, so the structure could be metal-rich, which leads to a rapid drop in metallicity beyond $r\sim80\,{\rm kpc}$ (i.e., its outer boundary).

In Figure \ref{vorbinfeh}, for the clean sample (bottom), the interface between metal-rich and metal-poor regions is located at $r\sim40\,$kpc, while for the total sample (top), three components, including a small structure with $\rm [Fe/H]\sim-1.85$ at $(R,|z|)\sim(26,30)\,{\rm kpc}$, the over-density structure with $\rm [Fe/H]\sim-1.70$ at low zenithal angles and the metal-rich Sgr stream with $\rm [Fe/H]\sim-1.65$, are distributed around $r\sim40$\,kpc.
Therefore, a quick growth of metallicity dispersion in the region of $30\,{\rm kpc}\,\textless\,r\,\textless\,40\,{\rm kpc}$ and a rapid drop beyond $r\sim40$\,kpc can be seen for two initial samples, respectively.
The extremely metal-rich structure at $(R,|z|)\sim(60,40)\,{\rm kpc}$ for the total sample (top) and the interface between the over-dense and metal-rich structure at low zenith angles and background halo beyond $r\sim80\,$kpc for two initial samples (top and bottom) lead to a growth in the metallicity dispersion around $r\sim80$\,kpc.
Although the situation described above will lead to the trend of metallicity with $r$ or $R-|z|$ as well as its dispersion with $r$ in Figures \ref{figr_feh_c} and \ref{vorbinfeh}, it could also be attributed to errors or some debris of undiscovered mergers in the outer halo.

\subsection{Shell-shaped structure}

In order to intuitively illustrate that the over-density structure does not belong to distant streams, we show the overdensities with $r\textgreater50\,{\rm kpc}$ found by the HDBSCAN clustering algorithm and the distant streams with heliocentric distances larger than 30\,kpc given in the Python package \textit{\textbf{galstream}} \citep{mateu2022galstreams} in Figures \ref{merger_pmlb} and \ref{mergerxyz}.
We found that the overdensities are mostly distributed at low latitudes,  as expected, and are different from the shapes of stellar streams. In addition, their proper motions, depicted by arrows, reveal a complex velocity field.

In order to focus on the over-density structure, we only show stars at low vertical distances in Figure \ref{shells_xyz}, and it can be found that the structure is shell-shaped.
We infer that the flattening density distribution in the range of $36-96\,{\rm kpc}$, the deviation of the major axis $(\gamma\,\textgreater\,0,\ \eta\sim 0)$ from the direction dominated by HAC and VOD, and the two breaks at $r_{\rm e}\sim48\,{\rm kpc}$ and $\sim93\,{\rm kpc}$ are attributed to the two shells in Figure \ref{shells_xyz}.
Some numerical simulations have shown that shells are made by mergers on radial orbits, while streams appear in more circular orbits \citep{johnston2008tracing,amorisco2015feathers,karademir2019outer,pop2018formation}.
As expected, the candidates shown in Figure \ref{shells_xyz} clearly exhibit a shell-shaped or ring-shaped structure at low zenithal angles, which is very similar to the case of small orbital angle (edge-on to the host galaxy) and impact angle \citep[radial orbit,][]{karademir2019outer}.
In this case, more stars will be distributed at lower zenithal angles forming a ring, while the accretion in the direction perpendicular to the host disk will make stars get large vertical velocities distributed at higher zenithal angles. 
Therefore, the flattening in the outer halo indicates that the accretion at the edge is more reasonable.
\citet{pop2018formation} have found that the shell-forming progenitors usually accreted with high stellar mass ratio ($\textgreater\,0.1$) on approximately radial orbits about 4-8 Gyr ago, and stripped about 1-4 Gyr.
Most of the resulting shells are phase-mixed if satellites are accreted too early, while satellites accreted recently will not have enough time to be stripped and form shells. Thus, a clear shell shown in Figure \ref{shells_xyz} indicates that its accretion time could be intermediate and its progenitor galaxy may be a massive galaxy, such as GSE \citep[$\rm M_{GE}/M_{MW}\sim0.24$, $8-11\rm Gyr\ ago$][]{helmi2018merger,belokurov2018co}.
Considering that a major merger contributes a large stellar mass to the centre of our host galaxy \citep{karademir2019outer}, and the high-eccentricity stars in the inner halo mostly belong to GSE except for a small amount of the Splash stars \citep{splash2020}, the shells in Figure \ref{shells_xyz} may be caused by the radial collision of GSE, similar to the shells found by \citet{MWshells2020} in the HAC and VOD regions.
\citet{distantechoes2022} recently found the apocentric shells of GSE debris, forming $\rm 60-90\,kpc$ counterparts to the $\rm 15-20\,kpc$ shells that are known to dominate the inner halo.
Therefore, we infer that the shells are likely to be the apocenter pile-ups of GSE in the outer halo, and could be associated with Outer Virgo Overdensity \citep{sesar2017100} and a coherent stream of retrograde stars encircling the Milky Way from 50-100\,kpc, in the same plane as the Sgr stream but moving in the opposite direction, found by \citet{distantechoes2022}.

\section{Conclusion and summary}\label{Conclusions and discussion}

In this paper, we use a sample of RRab stars released by Gaia DR3 to study the relationship between break radius and merger in the Milky Way, and to explore new mergers. 
We apply two methods (GFFM and the classical fitting method) to fit two initial samples with the new broken power law in order to probe all breaks, which includes an additional parameter indicating the broken scale. 
We found that $q$ in $36\,{\rm kpc}\,\textless\,r\,\textless\,96\,{\rm kpc}$ is much smaller than that in the inner halo according to the formula of its increasing with $r$ in the region of $r\,\textless\,36\,$kpc.
In addition, the major axis deviates significantly from the direction dominated by HAC and VOD in the region of $66-96\,{\rm kpc}$, and has significant uncertainty in the range of $36-66\,\rm kpc$.
Therefore, we attribute the breaks at both $r_{\rm e}\sim48\,$kpc and 93\,kpc to two unknown overdensities at low zenithal angles.
In the study of metallicity as a function of $r$, we found that some turning points of metallicity have corresponding breaks, such as $r\sim20$\,kpc and 30\,kpc, suggesting that there are likely to be the apocenter pile-ups of GSE at these locations.
We also analyze the metallicity distribution in $R-|z|$ space, and found that the metallicity in the region of over-density structure at low zenithal angles is richer than that of the outer background halo, indicating that the two overdensities are responsible for the breaks at $r_{\rm e}\sim48\,$kpc and 93\,kpc are likely to be metal-rich.

Finally, we apply HDBSCAN to select the candidates belonging to the two overdensities that lead to two breaks at $r_{\rm e}\sim48\,$kpc and 93\,kpc, and infer that two overdensities are shell-shaped or ring-shaped, which is consistent with previous numerical simulations of the mergers on high-eccentricity orbits.
We conclude that the two shells are the apocenter pile-ups of GSE in the outer halo, and are associated with Outer Virgo Overdensity \citep{sesar2017100} and a coherent stream of retrograde stars encircling the Milky Way from 50-100\,kpc, in the same plane as the Sgr stream but moving in the opposite direction, found by \citep{distantechoes2022}.
Due to lack of high-quality observation data, such as radial velocity and chemical abundance, we still do not know its dynamical properties. 
Therefore, here we only tentatively propose its existence, and with the releasement of more high-precision and long-distance stellar data, it can be studied in depth.

\section*{Acknowledgements}
We thank the referee for the insightful comments and suggestions, which have improved the paper significantly.
This work was supported by the National Natural Sciences Foundation of China (NSFC Nos: 11973042, 12090040, 12090044, 11973052 and 11873053).  
It was also supported by the Fundamental Research Funds for the Central Universities and the National Key R\&D Program of China No. 2019YFA0405501.  
This work has made use of data from the European Space Agency (ESA) mission
{\it Gaia} (\url{https://www.cosmos.esa.int/gaia}), processed by the {\it Gaia}
Data Processing and Analysis Consortium (DPAC,
\url{https://www.cosmos.esa.int/web/gaia/dpac/consortium}). Funding for the DPAC
has been provided by national institutions, in particular the institutions
participating in the {\it Gaia} Multilateral Agreement.

\section*{Data Availability}
The data supporting this article will be shared upon reasonable request sent to the corresponding authors.



\bibliographystyle{mnras}
\bibliography{yds.bbl} 

\begin{thebibliography}{}
\makeatletter
\relax
\def\mn@urlcharsother{\let\do\@makeother \do\$\do\&\do\#\do\^\do\_\do\%\do\~}
\def\mn@doi{\begingroup\mn@urlcharsother \@ifnextchar [ {\mn@doi@}
  {\mn@doi@[]}}
\def\mn@doi@[#1]#2{\def\@tempa{#1}\ifx\@tempa\@empty \href
  {http://dx.doi.org/#2} {doi:#2}\else \href {http://dx.doi.org/#2} {#1}\fi
  \endgroup}
\def\mn@eprint#1#2{\mn@eprint@#1:#2::\@nil}
\def\mn@eprint@arXiv#1{\href {http://arxiv.org/abs/#1} {{\tt arXiv:#1}}}
\def\mn@eprint@dblp#1{\href {http://dblp.uni-trier.de/rec/bibtex/#1.xml}
  {dblp:#1}}
\def\mn@eprint@#1:#2:#3:#4\@nil{\def\@tempa {#1}\def\@tempb {#2}\def\@tempc
  {#3}\ifx \@tempc \@empty \let \@tempc \@tempb \let \@tempb \@tempa \fi \ifx
  \@tempb \@empty \def\@tempb {arXiv}\fi \@ifundefined
  {mn@eprint@\@tempb}{\@tempb:\@tempc}{\expandafter \expandafter \csname
  mn@eprint@\@tempb\endcsname \expandafter{\@tempc}}}

\bibitem[\protect\citeauthoryear{{Amorisco}}{{Amorisco}}{2015}]{amorisco2015feathers}
{Amorisco} N.~C.,  2015, \mn@doi [\mnras] {10.1093/mnras/stv648}, \href
  {https://ui.adsabs.harvard.edu/abs/2015MNRAS.450..575A} {450, 575}

\bibitem[\protect\citeauthoryear{{Bellazzini}, {Ibata}, {Malhan}, {Martin},
  {Famaey}  \& {Thomas}}{{Bellazzini} et~al.}{2020}]{bellazzini2020globular}
{Bellazzini} M.,  {Ibata} R.,  {Malhan} K.,  {Martin} N.,  {Famaey} B.,
  {Thomas} G.,  2020, \mn@doi [\aap] {10.1051/0004-6361/202037621}, \href
  {https://ui.adsabs.harvard.edu/abs/2020A&A...636A.107B} {636, A107}

\bibitem[\protect\citeauthoryear{{Belokurov} et~al.,}{{Belokurov}
  et~al.}{2006}]{belokurov2006field}
{Belokurov} V.,  et~al., 2006, \mn@doi [\apjl] {10.1086/504797}, \href
  {https://ui.adsabs.harvard.edu/abs/2006ApJ...642L.137B} {642, L137}

\bibitem[\protect\citeauthoryear{{Belokurov} et~al.,}{{Belokurov}
  et~al.}{2007}]{belokurov2007}
{Belokurov} V.,  et~al., 2007, \mn@doi [\apjl] {10.1086/513144}, \href
  {https://ui.adsabs.harvard.edu/abs/2007ApJ...657L..89B} {657, L89}

\bibitem[\protect\citeauthoryear{{Belokurov} et~al.,}{{Belokurov}
  et~al.}{2014}]{belokurov2014precession}
{Belokurov} V.,  et~al., 2014, \mn@doi [\mnras] {10.1093/mnras/stt1862}, \href
  {https://ui.adsabs.harvard.edu/abs/2014MNRAS.437..116B} {437, 116}

\bibitem[\protect\citeauthoryear{{Belokurov}, {Erkal}, {Evans}, {Koposov}  \&
  {Deason}}{{Belokurov} et~al.}{2018}]{belokurov2018co}
{Belokurov} V.,  {Erkal} D.,  {Evans} N.~W.,  {Koposov} S.~E.,   {Deason}
  A.~J.,  2018, \mn@doi [\mnras] {10.1093/mnras/sty982}, \href
  {https://ui.adsabs.harvard.edu/abs/2018MNRAS.478..611B} {478, 611}

\bibitem[\protect\citeauthoryear{{Belokurov}, {Sanders}, {Fattahi}, {Smith},
  {Deason}, {Evans}  \& {Grand}}{{Belokurov} et~al.}{2020}]{splash2020}
{Belokurov} V.,  {Sanders} J.~L.,  {Fattahi} A.,  {Smith} M.~C.,  {Deason}
  A.~J.,  {Evans} N.~W.,   {Grand} R. J.~J.,  2020, \mn@doi [\mnras]
  {10.1093/mnras/staa876}, \href
  {https://ui.adsabs.harvard.edu/abs/2020MNRAS.494.3880B} {494, 3880}

\bibitem[\protect\citeauthoryear{{Bland-Hawthorn} \&
  {Gerhard}}{{Bland-Hawthorn} \& {Gerhard}}{2016}]{bland2016galaxy}
{Bland-Hawthorn} J.,  {Gerhard} O.,  2016, \mn@doi [\araa]
  {10.1146/annurev-astro-081915-023441}, \href
  {https://ui.adsabs.harvard.edu/abs/2016ARA&A..54..529B} {54, 529}

\bibitem[\protect\citeauthoryear{{Bovy}}{{Bovy}}{2015}]{bovy2015}
{Bovy} J.,  2015, \mn@doi [\apjs] {10.1088/0067-0049/216/2/29}, \href
  {https://ui.adsabs.harvard.edu/abs/2015ApJS..216...29B} {216, 29}

\bibitem[\protect\citeauthoryear{Campello, Moulavi  \& Sander}{Campello
  et~al.}{2013}]{campello2013density}
Campello R.~J.,  Moulavi D.,   Sander J.,  2013, in Pacific-Asia conference on
  knowledge discovery and data mining. pp 160--172

\bibitem[\protect\citeauthoryear{{Cappellari} \& {Copin}}{{Cappellari} \&
  {Copin}}{2003}]{vorbin2003}
{Cappellari} M.,  {Copin} Y.,  2003, \mn@doi [\mnras]
  {10.1046/j.1365-8711.2003.06541.x}, \href
  {https://ui.adsabs.harvard.edu/abs/2003MNRAS.342..345C} {342, 345}

\bibitem[\protect\citeauthoryear{{Chandra} et~al.,}{{Chandra}
  et~al.}{2022}]{distantechoes2022}
{Chandra} V.,  et~al., 2022, \mn@doi [arXiv e-prints]
  {10.48550/arXiv.2212.00806}, \href
  {https://ui.adsabs.harvard.edu/abs/2022arXiv221200806C} {p. arXiv:2212.00806}

\bibitem[\protect\citeauthoryear{{Clementini} et~al.,}{{Clementini}
  et~al.}{2022}]{clementini2022gaia}
{Clementini} G.,  et~al., 2022, arXiv e-prints, \href
  {https://ui.adsabs.harvard.edu/abs/2022arXiv220606278C} {p. arXiv:2206.06278}

\bibitem[\protect\citeauthoryear{{Cooper} et~al.,}{{Cooper}
  et~al.}{2011}]{cooper2011formation}
{Cooper} A.~P.,  et~al., 2011, \mn@doi [\apjl] {10.1088/2041-8205/743/1/L21},
  \href {https://ui.adsabs.harvard.edu/abs/2011ApJ...743L..21C} {743, L21}

\bibitem[\protect\citeauthoryear{{Das}, {Williams}  \& {Binney}}{{Das}
  et~al.}{2016}]{das2016characterizing}
{Das} P.,  {Williams} A.,   {Binney} J.,  2016, \mn@doi [\mnras]
  {10.1093/mnras/stw2167}, \href
  {https://ui.adsabs.harvard.edu/abs/2016MNRAS.463.3169D} {463, 3169}

\bibitem[\protect\citeauthoryear{{Das}, {Hawkins}  \& {Jofr{\'e}}}{{Das}
  et~al.}{2020}]{das2020ages}
{Das} P.,  {Hawkins} K.,   {Jofr{\'e}} P.,  2020, \mn@doi [\mnras]
  {10.1093/mnras/stz3537}, \href
  {https://ui.adsabs.harvard.edu/abs/2020MNRAS.493.5195D} {493, 5195}

\bibitem[\protect\citeauthoryear{{Deason}, {Belokurov}  \& {Evans}}{{Deason}
  et~al.}{2011}]{deason2011milky}
{Deason} A.~J.,  {Belokurov} V.,   {Evans} N.~W.,  2011, \mn@doi [\mnras]
  {10.1111/j.1365-2966.2011.19237.x}, \href
  {https://ui.adsabs.harvard.edu/abs/2011MNRAS.416.2903D} {416, 2903}

\bibitem[\protect\citeauthoryear{{Deason}, {Belokurov}, {Koposov}  \&
  {Rockosi}}{{Deason} et~al.}{2014}]{deason2014touching}
{Deason} A.~J.,  {Belokurov} V.,  {Koposov} S.~E.,   {Rockosi} C.~M.,  2014,
  \mn@doi [\apj] {10.1088/0004-637X/787/1/30}, \href
  {https://ui.adsabs.harvard.edu/abs/2014ApJ...787...30D} {787, 30}

\bibitem[\protect\citeauthoryear{{Deason}, {Belokurov}, {Koposov}  \&
  {Lancaster}}{{Deason} et~al.}{2018}]{deason2018apocenter}
{Deason} A.~J.,  {Belokurov} V.,  {Koposov} S.~E.,   {Lancaster} L.,  2018,
  \mn@doi [\apjl] {10.3847/2041-8213/aad0ee}, \href
  {https://ui.adsabs.harvard.edu/abs/2018ApJ...862L...1D} {862, L1}

\bibitem[\protect\citeauthoryear{{Donlon}, {Newberg}, {Sanderson}  \&
  {Widrow}}{{Donlon} et~al.}{2020}]{MWshells2020}
{Donlon} Thomas I.,  {Newberg} H.~J.,  {Sanderson} R.,   {Widrow} L.~M.,  2020,
  \mn@doi [\apj] {10.3847/1538-4357/abb5f6}, \href
  {https://ui.adsabs.harvard.edu/abs/2020ApJ...902..119D} {902, 119}

\bibitem[\protect\citeauthoryear{{Drake} et~al.,}{{Drake}
  et~al.}{2013}]{drake2013probing}
{Drake} A.~J.,  et~al., 2013, \mn@doi [\apj] {10.1088/0004-637X/763/1/32},
  \href {https://ui.adsabs.harvard.edu/abs/2013ApJ...763...32D} {763, 32}

\bibitem[\protect\citeauthoryear{{Forbes}}{{Forbes}}{2020}]{forbes2020reverse}
{Forbes} D.~A.,  2020, \mn@doi [\mnras] {10.1093/mnras/staa245}, \href
  {https://ui.adsabs.harvard.edu/abs/2020MNRAS.493..847F} {493, 847}

\bibitem[\protect\citeauthoryear{{Foreman-Mackey}, {Hogg}, {Lang}  \&
  {Goodman}}{{Foreman-Mackey} et~al.}{2013}]{foreman2013emcee}
{Foreman-Mackey} D.,  {Hogg} D.~W.,  {Lang} D.,   {Goodman} J.,  2013, \mn@doi
  [\pasp] {10.1086/670067}, \href
  {https://ui.adsabs.harvard.edu/abs/2013PASP..125..306F} {125, 306}

\bibitem[\protect\citeauthoryear{{GRAVITY Collaboration} et~al.,}{{GRAVITY
  Collaboration} et~al.}{2018}]{abuter2018detection}
{GRAVITY Collaboration} et~al., 2018, \mn@doi [\aap]
  {10.1051/0004-6361/201833718}, \href
  {https://ui.adsabs.harvard.edu/abs/2018A&A...615L..15G} {615, L15}

\bibitem[\protect\citeauthoryear{{Gaia Collaboration} et~al.,}{{Gaia
  Collaboration} et~al.}{2016}]{prusti2016gaia}
{Gaia Collaboration} et~al., 2016, \mn@doi [\aap]
  {10.1051/0004-6361/201629272}, \href
  {https://ui.adsabs.harvard.edu/abs/2016A&A...595A...1G} {595, A1}

\bibitem[\protect\citeauthoryear{{Gaia Collaboration} et~al.,}{{Gaia
  Collaboration} et~al.}{2022}]{vallenari2022gaia}
{Gaia Collaboration} et~al., 2022, arXiv e-prints, \href
  {https://ui.adsabs.harvard.edu/abs/2022arXiv220800211G} {p. arXiv:2208.00211}

\bibitem[\protect\citeauthoryear{{Goodman} \& {Weare}}{{Goodman} \&
  {Weare}}{2010}]{goodman2010ensemble}
{Goodman} J.,  {Weare} J.,  2010, \mn@doi [Communications in Applied
  Mathematics and Computational Science] {10.2140/camcos.2010.5.65}, \href
  {https://ui.adsabs.harvard.edu/abs/2010CAMCS...5...65G} {5, 65}

\bibitem[\protect\citeauthoryear{{Han} et~al.,}{{Han}
  et~al.}{2022}]{han2022stellar}
{Han} J.~J.,  et~al., 2022, \mn@doi [\aj] {10.3847/1538-3881/ac97e9}, \href
  {https://ui.adsabs.harvard.edu/abs/2022AJ....164..249H} {164, 249}

\bibitem[\protect\citeauthoryear{{Harris}}{{Harris}}{1996}]{harris1996catalog}
{Harris} W.~E.,  1996, \mn@doi [\aj] {10.1086/118116}, \href
  {https://ui.adsabs.harvard.edu/abs/1996AJ....112.1487H} {112, 1487}

\bibitem[\protect\citeauthoryear{{Helmi}, {White}, {de Zeeuw}  \&
  {Zhao}}{{Helmi} et~al.}{1999}]{helmi1999debris}
{Helmi} A.,  {White} S. D.~M.,  {de Zeeuw} P.~T.,   {Zhao} H.,  1999, \mn@doi
  [\nat] {10.1038/46980}, \href
  {https://ui.adsabs.harvard.edu/abs/1999Natur.402...53H} {402, 53}

\bibitem[\protect\citeauthoryear{{Helmi}, {Babusiaux}, {Koppelman}, {Massari},
  {Veljanoski}  \& {Brown}}{{Helmi} et~al.}{2018}]{helmi2018merger}
{Helmi} A.,  {Babusiaux} C.,  {Koppelman} H.~H.,  {Massari} D.,  {Veljanoski}
  J.,   {Brown} A. G.~A.,  2018, \mn@doi [\nat] {10.1038/s41586-018-0625-x},
  \href {https://ui.adsabs.harvard.edu/abs/2018Natur.563...85H} {563, 85}

\bibitem[\protect\citeauthoryear{{Hernitschek} et~al.,}{{Hernitschek}
  et~al.}{2017}]{hernitschek2017geometry}
{Hernitschek} N.,  et~al., 2017, \mn@doi [\apj] {10.3847/1538-4357/aa960c},
  \href {https://ui.adsabs.harvard.edu/abs/2017ApJ...850...96H} {850, 96}

\bibitem[\protect\citeauthoryear{{Hernitschek} et~al.,}{{Hernitschek}
  et~al.}{2018}]{hernitschek2018profile}
{Hernitschek} N.,  et~al., 2018, \mn@doi [\apj] {10.3847/1538-4357/aabfbb},
  \href {https://ui.adsabs.harvard.edu/abs/2018ApJ...859...31H} {859, 31}

\bibitem[\protect\citeauthoryear{{Ibata}, {Gilmore}  \& {Irwin}}{{Ibata}
  et~al.}{1994}]{ibata1994dwarf}
{Ibata} R.~A.,  {Gilmore} G.,   {Irwin} M.~J.,  1994, \mn@doi [\nat]
  {10.1038/370194a0}, \href
  {https://ui.adsabs.harvard.edu/abs/1994Natur.370..194I} {370, 194}

\bibitem[\protect\citeauthoryear{{Iorio} \& {Belokurov}}{{Iorio} \&
  {Belokurov}}{2019}]{iorio2019shape}
{Iorio} G.,  {Belokurov} V.,  2019, \mn@doi [\mnras] {10.1093/mnras/sty2806},
  \href {https://ui.adsabs.harvard.edu/abs/2019MNRAS.482.3868I} {482, 3868}

\bibitem[\protect\citeauthoryear{{Iorio} \& {Belokurov}}{{Iorio} \&
  {Belokurov}}{2021}]{iorio2021chemo}
{Iorio} G.,  {Belokurov} V.,  2021, \mn@doi [\mnras] {10.1093/mnras/stab005},
  \href {https://ui.adsabs.harvard.edu/abs/2021MNRAS.502.5686I} {502, 5686}

\bibitem[\protect\citeauthoryear{{Iorio}, {Belokurov}, {Erkal}, {Koposov},
  {Nipoti}  \& {Fraternali}}{{Iorio} et~al.}{2018}]{iorio2018first}
{Iorio} G.,  {Belokurov} V.,  {Erkal} D.,  {Koposov} S.~E.,  {Nipoti} C.,
  {Fraternali} F.,  2018, \mn@doi [\mnras] {10.1093/mnras/stx2819}, \href
  {https://ui.adsabs.harvard.edu/abs/2018MNRAS.474.2142I} {474, 2142}

\bibitem[\protect\citeauthoryear{{Johnston}, {Bullock}, {Sharma}, {Font},
  {Robertson}  \& {Leitner}}{{Johnston} et~al.}{2008}]{johnston2008tracing}
{Johnston} K.~V.,  {Bullock} J.~S.,  {Sharma} S.,  {Font} A.,  {Robertson}
  B.~E.,   {Leitner} S.~N.,  2008, \mn@doi [\apj] {10.1086/592228}, \href
  {https://ui.adsabs.harvard.edu/abs/2008ApJ...689..936J} {689, 936}

\bibitem[\protect\citeauthoryear{{Juri{\'c}} et~al.,}{{Juri{\'c}}
  et~al.}{2008}]{juric2008milky}
{Juri{\'c}} M.,  et~al., 2008, \mn@doi [\apj] {10.1086/523619}, \href
  {https://ui.adsabs.harvard.edu/abs/2008ApJ...673..864J} {673, 864}

\bibitem[\protect\citeauthoryear{{Karademir}, {Remus}, {Burkert}, {Dolag},
  {Hoffmann}, {Moster}, {Steinwandel}  \& {Zhang}}{{Karademir}
  et~al.}{2019}]{karademir2019outer}
{Karademir} G.~S.,  {Remus} R.-S.,  {Burkert} A.,  {Dolag} K.,  {Hoffmann}
  T.~L.,  {Moster} B.~P.,  {Steinwandel} U.~P.,   {Zhang} J.,  2019, \mn@doi
  [\mnras] {10.1093/mnras/stz1251}, \href
  {https://ui.adsabs.harvard.edu/abs/2019MNRAS.487..318K} {487, 318}

\bibitem[\protect\citeauthoryear{{Koppelman}, {Helmi}, {Massari},
  {Price-Whelan}  \& {Starkenburg}}{{Koppelman}
  et~al.}{2019}]{koppelman2019multiple}
{Koppelman} H.~H.,  {Helmi} A.,  {Massari} D.,  {Price-Whelan} A.~M.,
  {Starkenburg} T.~K.,  2019, \mn@doi [\aap] {10.1051/0004-6361/201936738},
  \href {https://ui.adsabs.harvard.edu/abs/2019A&A...631L...9K} {631, L9}

\bibitem[\protect\citeauthoryear{{Kruijssen} et~al.,}{{Kruijssen}
  et~al.}{2020}]{kruijssen2020kraken}
{Kruijssen} J.~M.~D.,  et~al., 2020, \mn@doi [\mnras] {10.1093/mnras/staa2452},
  \href {https://ui.adsabs.harvard.edu/abs/2020MNRAS.498.2472K} {498, 2472}

\bibitem[\protect\citeauthoryear{{Lepage}}{{Lepage}}{1978}]{lepage1978new}
{Lepage} G.~P.,  1978, \mn@doi [Journal of Computational Physics]
  {10.1016/0021-9991(78)90004-9}, \href
  {https://ui.adsabs.harvard.edu/abs/1978JCoPh..27..192L} {27, 192}

\bibitem[\protect\citeauthoryear{{Li} et~al.,}{{Li}
  et~al.}{2019}]{li2019detecting}
{Li} J.,  et~al., 2019, \mn@doi [\apj] {10.3847/1538-4357/ab09ef}, \href
  {https://ui.adsabs.harvard.edu/abs/2019ApJ...874..138L} {874, 138}

\bibitem[\protect\citeauthoryear{{Li}, {Huang}, {Liu}, {Beers}  \&
  {Zhang}}{{Li} et~al.}{2022}]{li2022photometric}
{Li} X.-Y.,  {Huang} Y.,  {Liu} G.-C.,  {Beers} T.~C.,   {Zhang} H.-W.,  2022,
  arXiv e-prints, \href {https://ui.adsabs.harvard.edu/abs/2022arXiv220607668L}
  {p. arXiv:2206.07668}

\bibitem[\protect\citeauthoryear{{Liu}}{{Liu}}{1991}]{liu1991synthetic}
{Liu} T.,  1991, \mn@doi [\pasp] {10.1086/132809}, \href
  {https://ui.adsabs.harvard.edu/abs/1991PASP..103..205L} {103, 205}

\bibitem[\protect\citeauthoryear{{Mackereth} \& {Bovy}}{{Mackereth} \&
  {Bovy}}{2020}]{mackereth2020weighing}
{Mackereth} J.~T.,  {Bovy} J.,  2020, \mn@doi [\mnras] {10.1093/mnras/staa047},
  \href {https://ui.adsabs.harvard.edu/abs/2020MNRAS.492.3631M} {492, 3631}

\bibitem[\protect\citeauthoryear{{Majewski}, {Skrutskie}, {Weinberg}  \&
  {Ostheimer}}{{Majewski} et~al.}{2003}]{majewski2003two}
{Majewski} S.~R.,  {Skrutskie} M.~F.,  {Weinberg} M.~D.,   {Ostheimer} J.~C.,
  2003, \mn@doi [\apj] {10.1086/379504}, \href
  {https://ui.adsabs.harvard.edu/abs/2003ApJ...599.1082M} {599, 1082}

\bibitem[\protect\citeauthoryear{{Malhan}, {Yuan}, {Ibata}, {Arentsen},
  {Bellazzini}  \& {Martin}}{{Malhan} et~al.}{2021}]{malhan2021evidence}
{Malhan} K.,  {Yuan} Z.,  {Ibata} R.~A.,  {Arentsen} A.,  {Bellazzini} M.,
  {Martin} N.~F.,  2021, \mn@doi [\apj] {10.3847/1538-4357/ac1675}, \href
  {https://ui.adsabs.harvard.edu/abs/2021ApJ...920...51M} {920, 51}

\bibitem[\protect\citeauthoryear{{Malhan} et~al.,}{{Malhan}
  et~al.}{2022}]{malhan2022global}
{Malhan} K.,  et~al., 2022, \mn@doi [\apj] {10.3847/1538-4357/ac4d2a}, \href
  {https://ui.adsabs.harvard.edu/abs/2022ApJ...926..107M} {926, 107}

\bibitem[\protect\citeauthoryear{{Mateu}}{{Mateu}}{2022}]{mateu2022galstreams}
{Mateu} C.,  2022, arXiv e-prints, \href
  {https://ui.adsabs.harvard.edu/abs/2022arXiv220410326M} {p. arXiv:2204.10326}

\bibitem[\protect\citeauthoryear{{Mateu}, {Read}  \& {Kawata}}{{Mateu}
  et~al.}{2018}]{mateu2018fourteen}
{Mateu} C.,  {Read} J.~I.,   {Kawata} D.,  2018, \mn@doi [\mnras]
  {10.1093/mnras/stx2937}, \href
  {https://ui.adsabs.harvard.edu/abs/2018MNRAS.474.4112M} {474, 4112}

\bibitem[\protect\citeauthoryear{{McInnes}, {Healy}  \& {Astels}}{{McInnes}
  et~al.}{2017}]{mcinnes2017hdbscan}
{McInnes} L.,  {Healy} J.,   {Astels} S.,  2017, \mn@doi [The Journal of Open
  Source Software] {10.21105/joss.00205}, \href
  {https://ui.adsabs.harvard.edu/abs/2017JOSS....2..205M} {2, 205}

\bibitem[\protect\citeauthoryear{{Miceli} et~al.,}{{Miceli}
  et~al.}{2008}]{miceli2008evidence}
{Miceli} A.,  et~al., 2008, \mn@doi [\apj] {10.1086/533484}, \href
  {https://ui.adsabs.harvard.edu/abs/2008ApJ...678..865M} {678, 865}

\bibitem[\protect\citeauthoryear{{Muraveva}, {Delgado}, {Clementini}, {Sarro}
  \& {Garofalo}}{{Muraveva} et~al.}{2018}]{muraveva2018rr}
{Muraveva} T.,  {Delgado} H.~E.,  {Clementini} G.,  {Sarro} L.~M.,   {Garofalo}
  A.,  2018, \mn@doi [\mnras] {10.1093/mnras/sty2241}, \href
  {https://ui.adsabs.harvard.edu/abs/2018MNRAS.481.1195M} {481, 1195}

\bibitem[\protect\citeauthoryear{{Myeong}, {Vasiliev}, {Iorio}, {Evans}  \&
  {Belokurov}}{{Myeong} et~al.}{2019}]{myeong2019evidence}
{Myeong} G.~C.,  {Vasiliev} E.,  {Iorio} G.,  {Evans} N.~W.,   {Belokurov} V.,
  2019, \mn@doi [\mnras] {10.1093/mnras/stz1770}, \href
  {https://ui.adsabs.harvard.edu/abs/2019MNRAS.488.1235M} {488, 1235}

\bibitem[\protect\citeauthoryear{{Naidu}, {Conroy}, {Bonaca}, {Johnson},
  {Ting}, {Caldwell}, {Zaritsky}  \& {Cargile}}{{Naidu}
  et~al.}{2020}]{naidu2020evidence}
{Naidu} R.~P.,  {Conroy} C.,  {Bonaca} A.,  {Johnson} B.~D.,  {Ting} Y.-S.,
  {Caldwell} N.,  {Zaritsky} D.,   {Cargile} P.~A.,  2020, \mn@doi [\apj]
  {10.3847/1538-4357/abaef4}, \href
  {https://ui.adsabs.harvard.edu/abs/2020ApJ...901...48N} {901, 48}

\bibitem[\protect\citeauthoryear{{Naidu} et~al.,}{{Naidu}
  et~al.}{2021}]{naidu2021reconstructing}
{Naidu} R.~P.,  et~al., 2021, \mn@doi [\apj] {10.3847/1538-4357/ac2d2d}, \href
  {https://ui.adsabs.harvard.edu/abs/2021ApJ...923...92N} {923, 92}

\bibitem[\protect\citeauthoryear{{Newberg}, {Yanny}  \& {Willett}}{{Newberg}
  et~al.}{2009}]{newberg2009discovery}
{Newberg} H.~J.,  {Yanny} B.,   {Willett} B.~A.,  2009, \mn@doi [\apjl]
  {10.1088/0004-637X/700/2/L61}, \href
  {https://ui.adsabs.harvard.edu/abs/2009ApJ...700L..61N} {700, L61}

\bibitem[\protect\citeauthoryear{{Perottoni}, {Limberg}, {Amarante}, {Rossi},
  {Queiroz}, {Santucci}, {P{\'e}rez-Villegas}  \& {Chiappini}}{{Perottoni}
  et~al.}{2022}]{perottoni2022}
{Perottoni} H.~D.,  {Limberg} G.,  {Amarante} J. A.~S.,  {Rossi} S.,  {Queiroz}
  A. B.~A.,  {Santucci} R.~M.,  {P{\'e}rez-Villegas} A.,   {Chiappini} C.,
  2022, \mn@doi [\apjl] {10.3847/2041-8213/ac88d6}, \href
  {https://ui.adsabs.harvard.edu/abs/2022ApJ...936L...2P} {936, L2}

\bibitem[\protect\citeauthoryear{{Pop}, {Pillepich}, {Amorisco}  \&
  {Hernquist}}{{Pop} et~al.}{2018}]{pop2018formation}
{Pop} A.-R.,  {Pillepich} A.,  {Amorisco} N.~C.,   {Hernquist} L.,  2018,
  \mn@doi [\mnras] {10.1093/mnras/sty1932}, \href
  {https://ui.adsabs.harvard.edu/abs/2018MNRAS.480.1715P} {480, 1715}

\bibitem[\protect\citeauthoryear{{Ramos}, {Mateu}, {Antoja}, {Helmi},
  {Castro-Ginard}, {Balbinot}  \& {Carrasco}}{{Ramos}
  et~al.}{2020}]{ramos2020full}
{Ramos} P.,  {Mateu} C.,  {Antoja} T.,  {Helmi} A.,  {Castro-Ginard} A.,
  {Balbinot} E.,   {Carrasco} J.~M.,  2020, \mn@doi [\aap]
  {10.1051/0004-6361/202037819}, \href
  {https://ui.adsabs.harvard.edu/abs/2020A&A...638A.104R} {638, A104}

\bibitem[\protect\citeauthoryear{{Sch{\"o}nrich}}{{Sch{\"o}nrich}}{2012}]{schonrich2012galactic}
{Sch{\"o}nrich} R.,  2012, \mn@doi [\mnras] {10.1111/j.1365-2966.2012.21631.x},
  \href {https://ui.adsabs.harvard.edu/abs/2012MNRAS.427..274S} {427, 274}

\bibitem[\protect\citeauthoryear{{Sch{\"o}nrich}, {Binney}  \&
  {Dehnen}}{{Sch{\"o}nrich} et~al.}{2010}]{schonrich2010local}
{Sch{\"o}nrich} R.,  {Binney} J.,   {Dehnen} W.,  2010, \mn@doi [\mnras]
  {10.1111/j.1365-2966.2010.16253.x}, \href
  {https://ui.adsabs.harvard.edu/abs/2010MNRAS.403.1829S} {403, 1829}

\bibitem[\protect\citeauthoryear{{Schwarz}}{{Schwarz}}{1978}]{Schwarz1978}
{Schwarz} G.,  1978, Annals of Statistics, \href
  {https://ui.adsabs.harvard.edu/abs/1978AnSta...6..461S} {6, 461}

\bibitem[\protect\citeauthoryear{{Sesar} et~al.,}{{Sesar}
  et~al.}{2013}]{sesar2013exploring}
{Sesar} B.,  et~al., 2013, \mn@doi [\aj] {10.1088/0004-6256/146/2/21}, \href
  {https://ui.adsabs.harvard.edu/abs/2013AJ....146...21S} {146, 21}

\bibitem[\protect\citeauthoryear{{Sesar}, {Hernitschek}, {Dierickx}, {Fardal}
  \& {Rix}}{{Sesar} et~al.}{2017}]{sesar2017100}
{Sesar} B.,  {Hernitschek} N.,  {Dierickx} M. I.~P.,  {Fardal} M.~A.,   {Rix}
  H.-W.,  2017, \mn@doi [\apjl] {10.3847/2041-8213/aa7c61}, \href
  {https://ui.adsabs.harvard.edu/abs/2017ApJ...844L...4S} {844, L4}

\bibitem[\protect\citeauthoryear{{Vasiliev} \& {Baumgardt}}{{Vasiliev} \&
  {Baumgardt}}{2021}]{vasiliev2021gaia}
{Vasiliev} E.,  {Baumgardt} H.,  2021, \mn@doi [\mnras]
  {10.1093/mnras/stab1475}, \href
  {https://ui.adsabs.harvard.edu/abs/2021MNRAS.505.5978V} {505, 5978}

\bibitem[\protect\citeauthoryear{{Vasiliev}, {Belokurov}  \&
  {Erkal}}{{Vasiliev} et~al.}{2021}]{Vasiliev2021tango}
{Vasiliev} E.,  {Belokurov} V.,   {Erkal} D.,  2021, \mn@doi [\mnras]
  {10.1093/mnras/staa3673}, \href
  {https://ui.adsabs.harvard.edu/abs/2021MNRAS.501.2279V} {501, 2279}

\bibitem[\protect\citeauthoryear{{Vivas} et~al.,}{{Vivas}
  et~al.}{2001}]{vivas2001}
{Vivas} A.~K.,  et~al., 2001, \mn@doi [\apjl] {10.1086/320915}, \href
  {https://ui.adsabs.harvard.edu/abs/2001ApJ...554L..33V} {554, L33}

\bibitem[\protect\citeauthoryear{Weiss \& Kulikowski}{Weiss \&
  Kulikowski}{1991}]{weiss1991computer}
Weiss S.~M.,  Kulikowski C.~A.,  1991, Computer systems that learn:
  classification and prediction methods from statistics, neural nets, machine
  learning, and expert systems.
Morgan Kaufmann Publishers Inc.

\bibitem[\protect\citeauthoryear{{Xue}, {Rix}, {Ma}, {Morrison}, {Bovy},
  {Sesar}  \& {Janesh}}{{Xue} et~al.}{2015}]{xue2015radial}
{Xue} X.-X.,  {Rix} H.-W.,  {Ma} Z.,  {Morrison} H.,  {Bovy} J.,  {Sesar} B.,
  {Janesh} W.,  2015, \mn@doi [\apj] {10.1088/0004-637X/809/2/144}, \href
  {https://ui.adsabs.harvard.edu/abs/2015ApJ...809..144X} {809, 144}

\bibitem[\protect\citeauthoryear{{Yuan}, {Smith}, {Xue}, {Li}, {Liu}, {Wang},
  {Li}  \& {Chang}}{{Yuan} et~al.}{2019}]{yuan2019revealing}
{Yuan} Z.,  {Smith} M.~C.,  {Xue} X.-X.,  {Li} J.,  {Liu} C.,  {Wang} Y.,  {Li}
  L.,   {Chang} J.,  2019, \mn@doi [\apj] {10.3847/1538-4357/ab2e09}, \href
  {https://ui.adsabs.harvard.edu/abs/2019ApJ...881..164Y} {881, 164}

\bibitem[\protect\citeauthoryear{{Yuan}, {Chang}, {Beers}  \& {Huang}}{{Yuan}
  et~al.}{2020}]{yuan2020low}
{Yuan} Z.,  {Chang} J.,  {Beers} T.~C.,   {Huang} Y.,  2020, \mn@doi [\apjl]
  {10.3847/2041-8213/aba49f}, \href
  {https://ui.adsabs.harvard.edu/abs/2020ApJ...898L..37Y} {898, L37}

\makeatother
\end{thebibliography}




\appendix

\section{the results from GFFM}

\begin{table*}
\scalebox{0.86}{
\begin{threeparttable}
    \caption{Fitting results from the GFMM with SBPL.}
    \label{table1}
	\begin{tabular}{lcccccc} 
		\hline
		range (kpc) & $r_{\rm break}$(kpc) & $n$ & $q$ & $\delta n$ & $\rm GF_{t}$ & $\rm GF_{max}$\\
		\hline
		(6,16) & $10_{-3}^{+4}$(6) & $1.6_{-0.5}^{+0.6}$(1,1-5,0.2) & $0.70_{-0.20}^{+0.10}$(0.75,0.1-1.0,0.05) & $0.6_{-0.4}^{+0.5}$(0.1,0.1-2.0,0.1) & 0.6 & 0.995 \\
         & $10_{-3}^{+4}$(7) & $1.6_{-0.6}^{+0.4}$(1,1-5,0.2) & $0.65_{-0.15}^{+0.15}$(0.75,0.1-1.0,0.05) & $0.6_{-0.3}^{+0.6}$(0.1,0.1-2.0,0.1) & 0.7 & \\
         & $10_{-3}^{+4}$(6) & $1.4_{-0.4}^{+0.6}$(1,1-5,0.2) & $0.65_{-0.15}^{+0.10}$(0.65,0.1-1.0,0.05) & $0.5_{-0.3}^{+0.5}$(0.1,0.1-2.0,0.1) & 0.8 & \\
         & $10_{-3}^{+4}$(6) & $1.4_{-0.4}^{+0.4}$(1,1-5,0.2) & $0.60_{-0.10}^{+0.10}$(0.65,0.1-1.0,0.05) & $0.5_{-0.3}^{+0.4}$(0.2,0.1-2.0,0.1) & 0.9 & \\
        (16,26) & $22_{-4}^{+3}$(26) & $2.5_{-0.5}^{+0.3}$(2.8,1.5-3,0.1) & $0.88_{-0.16}^{+0.10}$(1,0.6-1.0,0.01) & $0.5_{-0.2}^{+0.3}$(0.5,0.1-1.5,0.1) & 0.6 & 0.984\\
         & $23_{-5}^{+2}$(26) & $2.5_{-0.5}^{+0.3}$(2.8,1.5-3,0.1) & $0.88_{-0.18}^{+0.10}$(1,0.6-1.0,0.01) & $0.5_{-0.2}^{+0.3}$(0.4,0.1-1.5,0.1) & 0.7 & \\
         & $23_{-3}^{+3}$(26) & $2.5_{-0.4}^{+0.3}$(2.8,1.5-3,0.1) & $0.88_{-0.18}^{+0.10}$(1,0.6-1.0,0.01) & $0.4_{-0.1}^{+0.3}$(0.4,0.1-1.5,0.1) & 0.8 & \\
         & $24_{-2}^{+1}$(24) & $2.6_{-0.5}^{+0.3}$(2.8,1.5-3,0.1) & $0.84_{-0.18}^{+0.14}$(1,0.6-1.0,0.01) & $0.4_{-0.1}^{+0.2}$([0.3,0.4],0.1-1.5,0.1) & 0.9 & \\
        (26,36) & $31_{-3}^{+3}$(30) & $2.9_{-0.2}^{+0.3}$(3,2.5-3.5,0.1) & $0.86_{-0.11}^{+0.10}$(0.87,0.7-1.0,0.01) & $0.4_{-0.2}^{+0.2}$(0.4,0.1-1.0,0.1) & 0.6 & 0.994\\
         & $30_{-2}^{+4}$(30) & $2.8_{-0.3}^{+0.2}$(2.9,2.5-3.5,0.1) & $0.86_{-0.11}^{+0.10}$(0.84,0.7-1.0,0.01) & $0.4_{-0.2}^{+0.2}$(0.4,0.1-1.0,0.1) & 0.7 & \\
         & $30_{-2}^{+2}$(30) & $2.9_{-0.3}^{+0.2}$(2.9,2.5-3.5,0.1) & $0.85_{-0.10}^{+0.11}$(1,0.7-1.0,0.01) & $0.5_{-0.2}^{+0.1}$(0.4,0.1-1.0,0.1) & 0.8 & \\
         & $30_{-2}^{+1}$(30) & $2.8_{-0.2}^{+0.2}$(2.9,2.5-3.5,0.1) & $0.86_{-0.10}^{+0.10}$([0.78,0.88,0.89,0.97],0.7-1.0,0.01) & $0.5_{-0.2}^{+0.1}$(0.5,0.1-1.0,0.1) & 0.9 &  \\
        (36,46) & $43_{-1}^{+3}$(43) & $3.4_{-0.1}^{+0.0}$(3.3,3-4,0.1) & $0.80_{-0.07}^{+0.11}$([0.72,0.75,0.76],0.7-1.0,0.01) & $-0.3_{-0.0}^{+0.0}$(-0.3,-0.6\ -\ -0.1,0.1) & 0.6 & 0.870\\
         & $43_{-0}^{+2}$(43) & $3.4_{-0.1}^{+0.0}$([3.3,3.4],3-4,0.1) & $0.78_{-0.04}^{+0.08}$(0.75,0.7-1.0,0.01) & $-0.3_{-0.0}^{+0.0}$(-0.3,-0.6\ -\ -0.1,0.1) & 0.7 & \\
         & $43_{-0}^{+1}$(43) & $3.4_{-0.1}^{+0.0}$(3.4,3-4,0.1) & $0.85_{-0.11}^{+0.02}$([0.73,0.74,0.77,0.85,0.86,0.87,0.88],0.7-1.0,0.01) & $-0.3_{-0.0}^{+0.0}$(-0.3,-0.6\ -\ -0.1,0.1) & 0.8 & \\
        (46,56) & $52_{-3}^{+3}$(56) & $3.1_{-0.1}^{+0.2}$(3.1,3-4,0.1) & $0.88_{-0.10}^{+0.09}$(1,0.7-1.0,0.01) & $0.3_{-0.2}^{+0.1}$(0.3,0.1-0.6,0.1) & 0.6 & 0.988\\
         & $52_{-3}^{+3}$(56) & $3.1_{-0.1}^{+0.2}$(3.2,3-4,0.1) & $0.88_{-0.09}^{+0.08}$(0.95,0.7-1.0,0.01) & $0.3_{-0.1}^{+0.1}$(0.3,0.1-0.6,0.1) & 0.7 & \\
         & $52_{-3}^{+3}$([53,56]) & $3.1_{-0.1}^{+0.2}$(3.1,3-4,0.1) & $0.87_{-0.09}^{+0.08}$(0.95,0.7-1.0,0.01) & $0.3_{-0.1}^{+0.1}$(0.3,0.1-0.6,0.1) & 0.8 & \\
         & $52_{-2}^{+3}$(53) & $3.1_{-0.1}^{+0.1}$(3,3-4,0.1) & $0.85_{-0.08}^{+0.08}$([0.85,0.87],0.7-1.0,0.01) & $0.3_{-0.1}^{+0.1}$(0.3,0.1-0.6,0.1) & 0.9 & \\
        (56,76) & $60_{-4}^{+12}$(56) & $3.4_{-0.2}^{+0.1}$(3.4,3-4,0.1) & $0.89_{-0.07}^{+0.07}$(0.87,0.8-1.0,0.01) & $0.1_{-0.0}^{+0.2}$(0.1,0.1-0.6,0.1) & 0.6 & 0.935\\
         & $58_{-2}^{+10}$(56) & $3.3_{-0.2}^{+0.1}$(3.4,3-4,0.1) & $0.88_{-0.06}^{+0.05}$([0.88,0.89],0.8-1.0,0.01) & $0.1_{-0.0}^{+0.2}$(0.1,0.1-0.6,0.1) & 0.7 & \\
         & $57_{-1}^{+4}$(56) & $3.3_{-0.2}^{+0.1}$(3.4,3-4,0.1) & $0.89_{-0.08}^{+0.03}$(0.89,0.8-1.0,0.01) & $0.1_{-0.0}^{+0.2}$(0.1,0.1-0.6,0.1) & 0.8 & \\
         & $57_{-1}^{+2}$(56) & $3.3_{-0.1}^{+0.1}$(3.4,3-4,0.1) & $0.91_{-0.11}^{+0.01}$(0.8,0.8-1.0,0.01) & $0.1_{-0.0}^{+0.2}$(0.1,0.1-0.6,0.1) & 0.9 & \\
        (66,86) & $74.5_{-0.7}^{+0.7}$([74,75]) & $3.5_{-0.0}^{+0.0}$(3.5,3-4.5,0.1) & $0.85_{-0.01}^{+0.01}$(0.84,0.8-1.0,0.01) & $-0.1_{-0.0}^{+0.0}$(-0.1,-0.5\ -\ -0.1,0.1) & 0.6 & 0.644\\
        (76,96) & $90_{-11}^{+5}$(96) & $3.4_{-0.1}^{+0.1}$(3.4,3-4.5,0.1) & $0.90_{-0.07}^{+0.07}$(0.9,0.8-1.0,0.01) & $0.2_{-0.1}^{+0.1}$(0.1,0.1-0.6,0.1) & 0.6 & 0.945\\
         & $92_{-6}^{+3}$(96) & $3.4_{-0.0}^{+0.1}$(3.4,3-4.5,0.1) & $0.91_{-0.07}^{+0.07}$(1,0.8-1.0,0.01) & $0.2_{-0.1}^{+0.1}$(0.1,0.1-0.6,0.1) & 0.7 & \\
         & $93_{-4}^{+3}$(96) & $3.4_{-0.0}^{+0.1}$(3.5,3-4.5,0.1) & $0.91_{-0.06}^{+0.08}$(1,0.8-1.0,0.01) & $0.2_{-0.1}^{+0.1}$(0.1,0.1-0.6,0.1) & 0.8 & \\
         & $92_{-3}^{+3}$([90,91,92,96]) & $3.5_{-0.1}^{+0.0}$(3.5,3-4.5,0.1) & $0.96_{-0.10}^{+0.03}$([0.99,1],0.8-1.0,0.01) & $0.2_{-0.1}^{+0.1}$(0.1,0.1-0.6,0.1) & 0.9 & \\
        (96,116) & $107_{-5}^{+5}$(105) & $3.7_{-0.1}^{+0.2}$(3.8,3-4.5,0.1) & $0.90_{-0.07}^{+0.07}$(0.92,0.8-1.0,0.01) & $0.5_{-0.2}^{+0.1}$(0.6,0.1-0.6,0.1) & 0.6 & 0.959\\
         & $107_{-4}^{+4}$(106) & $3.7_{-0.1}^{+0.1}$(3.7,3-4.5,0.1) & $0.90_{-0.07}^{+0.07}$([0.87,0.88],0.8-1.0,0.01) & $0.5_{-0.1}^{+0.1}$(0.6,0.1-0.6,0.1) & 0.7 & \\
         & $106_{-3}^{+4}$(106) & $3.7_{-0.1}^{+0.1}$(3.7,3-4.5,0.1) & $0.90_{-0.07}^{+0.06}$(0.9,0.8-1.0,0.01) & $0.5_{-0.1}^{+0.1}$(0.6,0.1-0.6,0.1) & 0.8 & \\
         & $106_{-1}^{+3}$(106) & $3.7_{-0.1}^{+0.0}$(3.7,3-4.5,0.1) & $0.89_{-0.06}^{+0.06}$(0.87,0.8-1.0,0.01) & $0.6_{-0.1}^{+0.0}$(0.6,0.1-0.6,0.1) & 0.9 & \\
		\hline
	\end{tabular}
    \begin{tablenotes}
        \item From left to right, the interval in which we fit the data with SBPL, the break radius $r_{\rm break}$, the slope $n$, the flattening $q$, the increment of slope $\delta n$ at $r_{\rm break}$, the goodness-of-fit threshold ${\rm GF_t}$ and the largest goodness-of-fit $\rm GF_{max}$. 
        The value in parentheses in the second column is the mode. The quantities in parentheses from the third to fifth columns are the mode, the parameter range (the minimum and maximum) and the size of each grid, respectively. 
        The grid of $r_{\rm break}$ ($r_{\rm down}\,\textless\,r_{\rm break}\,\textless\,r_{\rm up}$) is 1\,kpc, and each point in the grid of $r_{\rm break}$ is integer. 
        Note that for the stars in 36\,kpc$\,\textless\,r\,\textless\,$56\,kpc and 66\,kpc$\,\textless\,r\,\textless\,$86\,kpc we have fitted SBPL between $\delta n=0.1$ and 0.6 but the goodness-of-fit is much lower than 0.6.
    \end{tablenotes}
\end{threeparttable}}
\end{table*}

\begin{table*}
\scalebox{1}{
\begin{threeparttable}
    \caption{Fitting results from the GFMM with SPL.}
    \label{table2}
	\begin{tabular}{lccccc} 
		\hline
		range (kpc) & $n$ & $q$ & $\rm GF_{t}$ & $\rm GF_{max}$ & $\Delta {\rm GF_{max}}$\\
		\hline
		(6,16) & $2.0_{-0.6}^{+0.7}$(1-5,0.2) & $0.63_{-0.23}^{+0.18}$(0.1-1.0,0.05) & 0.6 & 0.993 & -0.002 \\
         & $2.0_{-0.6}^{+0.6}$ & $0.60_{-0.20}^{+0.15}$ & 0.7 &  \\
         & $1.9_{-0.6}^{+0.5}$ & $0.58_{-0.35}^{+0.72}$ & 0.8 &  \\
         & $2.0_{-0.4}^{+0.2}$ & $0.55_{-0.10}^{+0.15}$ & 0.9 &  \\
        (16,26) & $2.9_{-0.0}^{+0.0}$(1.5-4.5,0.1) & $0.97_{-0.01}^{+0.02}$(0.6-1.0,0.01) & 0.6 & 0.649 & -0.335\\
        (6,26) & $2.6_{-0.2}^{+0.2}$(1.0-4.0,0.1) & $0.78_{-0.10}^{+0.09}$(0.5-1.0,0.01) & 0.6 & 0.811 & -0.183\\
         & $2.6_{-0.1}^{+0.2}$ & $0.78_{-0.07}^{+0.06}$ & 0.7  \\
         & $2.6_{-0.0}^{+0.1}$ & $0.79_{-0.02}^{+0.03}$ & 0.8  \\
        (26,36) & $3.3_{-0.0}^{+0.0}$(1.5-4.5,0.1) & $0.96_{-0.02}^{+0.02}$(0.6-1.0,0.01) & 0.6 & 0.628 & -0.366\\
        (36,46) & $3.0_{-0.0}^{+0.0}$(1.5-4.5,0.1) & $0.63_{-0.00}^{+0.00}$(0.6-1.0,0.01) & -3.016 & -3.016 & -3.886\\
        (46,56) & $3.3_{-0.0}^{+0.1}$(1.5-4.5,0.1) & $0.88_{-0.05}^{+0.10}$(0.6-1.0,0.01) & 0.5 & 0.577 & -0.411\\
        (56,76) & $3.4_{-0.1}^{+0.1}$(1.5-4.5,0.1) & $0.81_{-0.13}^{+0.13}$(0.6-1.0,0.01) & 0.6 & 0.914 & -0.021\\
         & $3.5_{-0.1}^{+0.1}$ & $0.81_{-0.12}^{+0.12}$ & 0.7 & \\
         & $3.5_{-0.1}^{+0.1}$ & $0.82_{-0.12}^{+0.12}$ & 0.8 & \\
         & $3.5_{-0.1}^{+0.0}$ & $0.93_{-0.07}^{+0.01}$ & 0.9 & \\
        (46,76) & $3.4_{-0.1}^{+0.1}$(1.5-4.5,0.1) & $0.83_{-0.13}^{+0.13}$(0.6-1.0,0.01) & 0.6 & 0.746 & -0.242\\
         & $3.4_{-0.0}^{+0.1}$ & $0.92_{-0.04}^{+0.06}$ & 0.7 \\
        (66,86) & $3.3_{-0.0}^{+0.0}$(1.5-4.5,0.1) & $0.69_{-0.00}^{+0.00}$(0.6-1.0,0.01) & 0.287 & 0.287 & -0.357\\
        (76,96) & $3.5_{-0.1}^{+0.1}$(1.5-4.5,0.1) & $0.86_{-0.09}^{+0.09}$(0.6-1.0,0.01) & 0.6 & 0.673 & -0.272\\
        (96,116) & $4.0_{-0.0}^{+0.0}$(1.5-4.5,0.1) & $0.97_{-0.00}^{+0.00}$(0.6-1.0,0.01) & 0.281 & 0.281 & -0.678\\
        
		\hline
	\end{tabular}
    \begin{tablenotes}
        \item From left to right, the interval in which we fit the data with SPL, the slope $n$, the flattening $q$, the goodness-of-fit threshold ${\rm GF_t}$, the largest goodness-of-fit $\rm GF_{max}$ and the difference between the $\rm GF_{max}$ of SPL and that of SBPL ($\rm \Delta GF_{max}$).
        The quantities in parentheses from the second to third columns are the parameter range (the minimum and maximum) and the size of each grid, respectively. 
    \end{tablenotes}
\end{threeparttable}}
\end{table*}

In Table \ref{table1}, we give the results obtained by fitting SBPL with GFFM in each interval.
However, the most frequent value is not very informative because the probability distribution of its neighborhood is so larger and not very different for some parameters, and there are even multiple values in some cases, that is, the most frequent values of some parameters are less prominent. Even so, we give all modes.
It is worth noting that the range of $\delta n$ in some groups is $(-0.5, -0.1)$ or $(-0.6, -0.1),$ because the GFs of results we obtain in the positive range are much smaller than 0.6.
In Table \ref{table2}, we also list the results obtained by fitting SPL to the data with GFFM in each bin.

\section{the results from classical fitting method}

In Table \ref{results from the classical fitting method} we give the results obtained by the classical fitting method. 
The probability distribution for each parameter in the region of $36\,{\rm kpc}\,\textless\,r\,\textless\,66\,{\rm kpc}$ is shown in Figures \ref{33_66_0break}-\ref{33_66_2break}.
\begin{table*}
\centering
\scalebox{0.8}{
\begin{threeparttable}
    \caption{\centering Fitting results from the classical fitting method.}
    \label{results from the classical fitting method}
	\begin{tabular}{lccccccccccccr} 
		\hline
		range (kpc) & $r_{\rm break1}~({\rm kpc})$ & $r_{\rm break2}~({\rm kpc})$ & $n$ & $q$ & $\delta n_1$ & $\delta n_2$ & $a_1$ & $a_2$ & $p$ & $\gamma$ & $\eta$ & $\beta$ & $\rm \Delta BIC$\\
		\hline
        $(6,26)$ &  &  & $2.64^{+0.11}_{-0.11}$ & $0.77^{+0.03}_{-0.03}$ & & & & & $1.45^{+0.07}_{-0.05}$ & $-0.21^{+0.08}_{-0.07}$ & $0.10^{+0.03}_{-0.03}$ & $0.22^{+0.08}_{-0.07}$ & 0 \\
        
		 &$22.99^{+0.91}_{-0.87}$ &  & $2.34^{+0.13}_{-0.10}$ & $0.81^{+0.02}_{-0.02}$ & $0.52^{+0.04}_{-0.05}$ & & $0.01^{+0.13}_{-0.14}$ & & $1.32^{+0.05}_{-0.03}$ & $-0.22^{+0.08}_{-0.07}$ & $0.10^{+0.03}_{-0.03}$ & $0.25^{+0.06}_{-0.08}$ & -86.11\\
    
        $(26,36)$ &  &  & $4.48^{+0.34}_{-0.37}$ & $0.89^{+0.02}_{-0.02}$ &  &  &  &  & $1.26^{+0.04}_{-0.04}$ & $-0.27^{+0.03}_{-0.05}$ & $0.17^{+0.03}_{-0.04}$ & $0.64^{+0.08}_{-0.09}$ & 0\\
        
         & $27.18^{+1.10}_{-0.67}$ &  & $2.94^{+0.14}_{-0.13}$ & $0.89^{+0.02}_{-0.02}$ & $0.53^{+0.11}_{-0.07}$ &  & $-0.08^{+0.17}_{-0.19}$ &  & $1.20^{+0.02}_{-0.02}$ & $-0.27^{+0.05}_{-0.05}$ & $0.14^{+0.03}_{-0.04}$ & $0.52^{+0.10}_{-0.10}$ & -12.90 \\
        
        $(36,66)$ & & & $3.29^{+0.16}_{-0.15}$ & $0.68^{+0.02}_{-0.03}$ & & & & & $1.09^{+0.04}_{-0.04}$ & $-0.22^{+0.20}_{-0.29}$ & $0.11^{+0.06}_{-0.04}$ & $0.15^{+0.06}_{-0.06}$ & 0 \\
        
         & $46.83^{+5.76}_{-4.25}$ & & $2.75^{+0.34}_{-0.34}$ & $0.69^{+0.03}_{-0.02}$ & $0.21^{+0.13}_{-0.15}$ & & $-0.87^{+0.41}_{-0.35}$ & & $1.07^{+0.05}_{-0.03}$ & $0.42^{+0.37}_{-0.42}$ & $0.01^{+0.08}_{-0.09}$ & $0.21^{+0.04}_{-0.04}$ & 22.03 \\

         & $45.25^{+2.51}_{-3.27}$ & $48.05^{+6.82}_{-3.49}$ & $3.42^{+0.22}_{-0.26}$ & $0.69^{+0.03}_{-0.03}$ & $-0.42^{+0.19}_{-0.13}$ & $0.31^{+0.14}_{-0.16}$ & $-0.75^{+0.62}_{-0.48}$ & $-0.66^{+0.88}_{-0.64}$ & $1.09^{+0.05}_{-0.04}$ & $-0.41^{+0.53}_{-0.50}$ & $0.16^{+0.06}_{-0.09}$ & $0.13^{+0.09}_{-0.10}$ & 71.10 \\
        
        $(66,96)$ &  &  & $3.92^{+0.18}_{-0.23}$ & $0.73^{+0.03}_{-0.04}$ &  &  &  &  & $1.28^{+0.04}_{-0.04}$ & $0.04^{+0.07}_{-0.06}$ & $0.02^{+0.02}_{-0.02}$ & $0.10^{+0.03}_{-0.04}$ & 0\\

         & $92.99^{+1.65}_{-2.84}$ &  & $4.71^{+0.42}_{-0.33}$ & $0.68^{+0.03}_{-0.02}$ & $-0.22^{+0.07}_{-0.07}$ &  & $0.44^{+0.50}_{-0.46}$ &  & $1.28^{+0.04}_{-0.04}$ & $0.12^{+0.05}_{-0.06}$ & $0.01^{+0.01}_{-0.02}$ & $0.09^{+0.03}_{-0.03}$ & -33.62\\
        
        $(96,116)$ & $103.77^{+3.76}_{-4.03}$ &  & $4.49^{+1.10}_{-1.80}$ & $0.92^{+0.02}_{-0.02}$ & $1.52^{+0.33}_{-0.47}$ &  & $-1.02^{+0.22}_{-0.16}$ &  & $1.04^{+0.02}_{-0.02}$ & $-0.40^{+0.17}_{-0.15}$ & $-0.30^{+0.19}_{-0.20}$ & $1.00^{+0.13}_{-0.14}$ & 0\\
		\hline
	\end{tabular}
    \begin{tablenotes}
        \item Note that $\rm \Delta BIC$ is the difference between the BIC values of SBPL or DBPL and that of SPL in each interval.
    \end{tablenotes}
\end{threeparttable}}
\end{table*}


\bsp	
\label{lastpage}
\end{document}